\documentclass[]{elsarticle}

\usepackage{graphicx} % Required for inserting images
\usepackage{subcaption}
\usepackage{caption}
\usepackage{geometry}
\usepackage{blindtext}
\usepackage{xcolor}
\usepackage{float}
\usepackage{amsmath}

\usepackage{amssymb}
\usepackage{tcolorbox}
\usepackage{amsfonts}
\usepackage{hyperref}
\usepackage{algorithmic} 
\usepackage{graphicx}
\usepackage{textcomp}
\usepackage{xcolor}
\usepackage{xspace}
\usepackage{multirow}
\usepackage{booktabs}
\usepackage{enumerate}
\usepackage{soul}
\usepackage{balance}
\usepackage{tcolorbox}
\usepackage{array}
\usepackage{multirow}
\usepackage{caption}
\usepackage{subcaption}
\usepackage[inline]{enumitem}
\usepackage[T1]{fontenc}
\usepackage{lipsum}
\usepackage{graphicx}
\usepackage{rotating}
\usepackage{microtype}
\usepackage{color}
\usepackage{float}
\usepackage{tcolorbox}
\newcommand{\citeTodo}[1]{{[??]}}

\newcommand{\ie}{i.e.,\xspace}

\newcommand{\eg}{e.g.,\xspace}

\newcommand{\COMMENT}[1]{}
\newboolean{showcomments}
\setboolean{showcomments}{true}         
\ifthenelse{\boolean{showcomments}}
  {\newcommand{\nb}[2]{
  \fbox{\bfseries\sffamily\scriptsize#1}
     {\sf\small$\blacktriangleright$\textit{{#2}}$\blacktriangleleft$}
   }
  }
  {\newcommand{\nb}[2]{}
   
  }

\newcommand{\sensodat}[0]{SensoDat\xspace}
\newcommand{\beamng}[0]{BeamNG.tech\xspace}

\NewTColorBox{custombox}{o}{
    colback=gray!10,
	colframe=gray!10,
	left=1.0mm,
	right=1.0mm,
	top=1.0mm,
	bottom=1.0mm,
	fonttitle=\bfseries,
	arc=0mm,
	leftrule=0mm,
	rightrule=0mm,
	toprule=0mm,
	bottomrule=0mm,
	notitle,
	before=\par\medskip\noindent,
    IfValueTF={#1}{before upper={\textbf{#1: } }}{},
    parbox=false,
}

\title{The Role of Road Features and Vehicle Dynamics in Cost-Effective Autonomous Vehicles Safety Testing: Insights from Instance Space Analysis
}

\author[1]{Victor Crespo-Rodriguez}
\ead{victor.cresporodriguez@monash.edu}

\author[3]{Christian Birchler}
\ead{christian.birchler@{zhaw,unibe}.ch}

\author[2]{Neelofar}
\ead{neelofar.neelofar@monash.edu}

\author[1]{Aldeida Aleti}
\ead{aldeida.aleti@monash.edu}

\author[3,4]{Sebastiano Panichella}
\ead{sebastiano.panichella@unibe.ch}

\affiliation[1]{
    organization={Monash University},
    country={Australia}}

\affiliation[2]{
    organization={RMIT University},
    country={Australia}
}
\affiliation[3]{
    organization={University of Bern},
    country={Switzerland}
}
\affiliation[4]{
    organization={Italian Institute of Artificial Intelligence for Industry (AI4I)},
    country={Italy}
}

\begin{document}

\begin{abstract}

\textbf{Context:} Simulation-based testing is a cost-efficient alternative to field testing for Autonomous Vehicles (AVs), but generating safety-critical test cases is challenging due to the vast search space. Prior work has studied static (road features) and dynamic (AV behavior) features of test scenarios separately, but their inter-dependencies are underexplored.

\textbf{Objective:} In this paper, we describe an empirical to analyze how static and dynamic features of test scenarios, and their inter-dependencies, influence AV test scenario outcomes.

\textbf{Method:} This study proposes an integrated approach using Instance Space Analysis (ISA) to evaluate both types of features, identify key influences on AV safety, and predict test outcomes without execution.

\textbf{Results:} Our study identifies critical features affecting test outcomes (effective/ineffective, depending on whether it leads to a safety-critical condition). Results show that combining static and dynamic features improves prediction accuracy, confirmed by models trained on both feature types outperforming models trained with only one type of feature.

\textbf{Conclusion:} The interplay of static and dynamic features enhances fault detection in AV testing. This research underscores the importance of integrating both types of features to create more effective testing frameworks for autonomous systems. Key contributions include: (1) a unified framework for AV safety assessment, (2) identification of influential features using ISA, and (3) efficient test outcome prediction for optimized regression testing.

% Simulation-based testing has emerged as a cost-efficient and replicable alternative to field testing for Autonomous Vehicles (AVs). However, generating numerous test cases to identify safety violations re0mains challenging due to the vast search space of possible configurations. While prior research has separately examined static (road features) and dynamic (AV behavior) factors influencing safety-critical test cases, their inter-dependencies are underexplored. This study proposes an integrated approach using Instance Space Analysis (ISA) to analyze both factors, identify key influences on AV safety, and develop cost-effective strategies to predict test outcomes without execution.\\
% Key contributions include: 1) a unified framework combining static and dynamic factors for AV safety assessment, 2) identification of influential features using ISA to improve test case effectiveness, and 3) cost-effective methods to predict test outcomes, optimizing regression testing. The findings highlight the critical relationship between static and dynamic factors in enhancing fault detection, ultimately advancing the development of safer and more reliable AVs. This research underscores the importance of integrating both factors to create more effective testing frameworks for autonomous systems.

\end{abstract}

% \begin{document}

\maketitle

\section{Introduction}
\label{sec:introduction}

Autonomous Vehicles (AVs) are employed across diverse domains \cite{HildebrandtE21,aerialist,anybotics2025simulation,GambiJRZ22,NguyenHG21}, yet ensuring their operational safety remains a critical challenge ~\cite{ZampettiKPP22, zhong2021survey, 10.1145/3377811.3380353,superialist}. In safety-critical applications, such as self-driving cars \cite{DBLP:journals/ethicsit/Stilgoe21}, failures can pose serious risks to human lives and the environment \cite{TheGuardian-2018}.
Consequently, rigorous testing is essential before real-world deployment to validate AV safety under practical conditions \cite{ZampettiKPP22, parekh2022review,DBLP:conf/icst/Jahangirova0T21}. However, real-world testing is prohibitively expensive, making it infeasible to verify AV reliability and safety across the millions of kilometers required for robust validation \cite{DBLP:conf/icst/KhatiriPT23,anybotics2025simulation}.

To address these challenges, recent research has proposed simulation-based testing strategies for AVs, particularly focusing on automated test generation for efficient evaluation  ~\cite{DBLP:conf/iros/ParraO0H23, DBLP:conf/icse/BiagiolaKPR23, GambiJRZ22}. Compared to real-world testing, simulation offers key advantages: it is highly reproducible, scalable, and cost-effective ~\cite{HildebrandtE21,birchler2021automated,birchler2024roadmapsimulationbasedtestingautonomous}. However, simulating AV behavior requires the creation of a vast number of driving scenarios (test cases). To maximize efficiency, test generation must prioritize safety-critical scenarios—those most likely to expose AV failures in simulation before they manifest in real-world conditions \cite{SDCScissor,birchler2024roadmapsimulationbasedtestingautonomous}. 

Existing approaches rely on safety oracles (e.g., predefined metrics) to classify system behavior as safe or unsafe ~\cite{DBLP:conf/sbst/PanichellaGZR21,GambiJRZ22,DBLP:journals/ese/BirchlerKBGP23}. Among these, the Out-of-Bound (OOB) metric, which evaluates an AV’s lateral position, is widely adopted for AV testing ~\cite{GambiJRZ22,NguyenHG21, DBLP:conf/icst/Jahangirova0T21}.
Despite its prevalence, generating safety-critical scenarios using OOB remains highly complex. This stems from three key challenges: (1) Dynamic environments (e.g., traffic, weather), (2) Numerous influencing factors (e.g., sensor noise, control algorithms),  and (3) An enormous search space of possible test configurations ~\cite{GambiJRZ22,NguyenHG21,DBLP:conf/icst/Jahangirova0T21} .

Recent research has investigated factors that characterize safety-critical test cases for autonomous vehicles (AVs), primarily examining either \textbf{static features} (\eg road geometry) \cite{DBLP:conf/icst/Jahangirova0T21,birchler2021automated,DBLP:journals/ese/BirchlerKBGP23,VR-study} or \textbf{dynamic features} (\eg AV behavioral properties) \cite{DBLP:conf/icst/Jahangirova0T21,10.1145/3377811.3380353,VR-study}. Static features encompass invariant road characteristics such as lane structure, curvature, intersections, and signage \cite{DBLP:journals/ese/BirchlerKBGP23,VR-study}, which remain unchanged during test execution. In contrast, dynamic features describe the AV operational state (\eg speed, steering angle, acceleration) and external variables like weather and lighting conditions \cite{DBLP:conf/icst/Jahangirova0T21,10.1145/3377811.3380353,VR-study}.

Despite these advances, a critical gap persists: no study has systematically analyzed how static and dynamic features \textit{interact} to influence test outcomes, particularly in determining safety violations. This oversight limits the comprehensiveness of simulation-based testing, as real-world driving scenarios inherently involve synergies between road infrastructure and vehicle dynamics. Furthermore, the relationship between static features (e.g., sharp curves) and their impact on dynamic AV behavior (e.g., speed adjustments) remains underexplored. Addressing these gaps would enable more robust test generation strategies, improving fault detection across diverse operational conditions and enhancing the reliability of AV safety assessments.

This study aims to systematically characterize the key factors influencing AV behavior in safety-critical test scenarios, with particular focus on the combined effects of both static and dynamic features \cite{DBLP:conf/icst/Jahangirova0T21, 10.1145/3377811.3380353,VR-study}. While prior research has examined these feature categories independently, our work represents the first comprehensive investigation into their joint influence on test case outcomes (\ie safety outcome).

The relationship between these features is multifaceted: not only do they vary across test scenarios, but static elements such as road geometry and intersection design directly affect dynamic AV responses including speed adjustments and steering maneuvers. This interdependence underscores the critical need for holistic analysis in test generation to ensure AV system reliability.

Nevertheless, defining the search space based on all features and their interactions significantly increases its complexity. Consequently, exploring this entire space becomes impractical due to the substantial time and resources needed. A more efficient strategy would ideally involve a focused investigation into the features that most strongly influence the effectiveness of test cases, particularly in detecting AV behavior faults (\ie unsafe behaviors). Prioritizing these key features allows for a more focused and efficient search. While recent work \cite{neelofar2024identifying, cresporodriguez2024isa} has explored feature selection methodologies (\eg Instance Space Analysis), these approaches maintain the artificial separation between static and dynamic characteristics, missing their synergistic effects.

In contrast to the aforementioned studies, which often treat static and dynamic features in isolation, our research introduces a novel integrated approach that concurrently analyzes both types of features within simulation environments. Unlike previous work, our methodology not only identifies the most influential factors affecting AV test cases but also delves into the inter-dependencies between static and dynamic features, offering a more comprehensive understanding of their combined impact. This represents an important, complementary focus from traditional approaches, providing a comprehensive framework that bridges the gap between road-based and AV-based analyses.
To gain the necessary insights into the impact of dynamic and static features on the simulation-based testing of AVs, in our study we investigate the following research questions:

\textbf{RQ1:} \textit{What are the most significant dynamic and static features of effective test cases?} 
Understanding road geometry (static features) and AV operation (dynamic features) during testing provides valuable insights into what makes a testing technique effective—defined by its ability to detect system faults. In this research question, we experiment with Instance Space Analysis (ISA) to identify the static and dynamic features, or test parameters, that contribute to the effectiveness of AV testing (i.e., to the identification of failing test cases). Studying static features helps to understand the role of road design in creating effective test cases, while dynamic features capture AV behavior leading to fault detection. The analysis with the ISA can visually illustrate and provide insights on how static and dynamic features (co-)influence test effectiveness.

\textbf{RQ2:} \textit{ How do static road features influence the operation of the AV under test, referred also as dynamic features?} 
The features of an effective test case reveal the road characteristics and AV operations present in effective scenarios. Understanding the actions taken by the AV during these scenarios can help identify the components or actuator systems that have the greatest influence on the outcome. Similarly, knowing the road characteristics where effective test cases occurred can provide insights into the AV's behavior under test. This research question focuses on gathering such insights, which are relevant for practitioners to focus their testing efforts with effective test cases that uncover relevant AV faults. 

\textbf{RQ3:} \textit{Can we effectively predict test case outcomes using both dynamic and static features?}
In this research question, we investigate whether the key features/factors studied in RQ1-2 can be used to train machine learning classifiers that allow us to predict test case outcomes (i.e., their safety) without executing the actual test cases. This investigation is conducted to streamline the process of prioritizing safety-critical test cases that are more likely to result in collisions, improving testing efficiency.

To address these questions, we extract dynamic and static features from AV testing suites with known test case outcomes \cite{DBLP:journals/corr/abs-2401-09808}. These features include road structure and the AV’s internal states during execution. Using them, we create an instance space that shows their influence on scenario outcomes. Scenarios are labeled as effective if the vehicle drifts from the lane center, causing an OOB incident ~\cite{GambiJRZ22,VR-study}; otherwise, they are considered ineffective.
We select the most impactful features to generate an instance space that visually highlights their effect on outcomes. 
 
To evaluate feature relevance more comprehensively, we trained multiple machine learning classifiers to predict test case outcomes using three distinct feature configurations: (1) static road features alone, (2) dynamic AV features alone, and (3) both feature sets combined. Our analysis revealed that the highest classification performance, measured across precision, recall, and F1-score metrics, was consistently achieved when integrating both static and dynamic features.

In summary, the contributions of this paper are:
\begin{enumerate}
    \item An integrated analysis of both static and dynamic features in simulation environments, which not only identifies the most influential factors impacting the effectiveness of AV test cases but also investigates their inter-dependencies.
    \item An examination of the interplay between static and dynamic features, revealing how road conditions and AV behaviors influence one another, thereby enhancing the understanding of factors contributing to safety incidents during AV testing.
    \item A comparative assessment of the importance of static and dynamic features by training machine learning classifiers on these features to predict test case outcomes prior to execution, providing insights into their relative predictive power.
\end{enumerate}

Our results demonstrate that an approach based on ISA allows practitioners to generate effective test cases using a smaller set of features, reducing testing time and resource consumption while maintaining test suite performance.
By identifying the internal components activated during successful test cases, practitioners can strategically prioritize the generation and execution of test cases, enabling targeted fault detection in specific AV components linked to both static and dynamic features. This prioritization not only enhances testing efficiency but also reduces resource expenditure. Furthermore, the capability to predict test case outcomes without actual execution allows for an efficient selection and prioritization of test scenarios, optimizing the testing process for AVs. These findings underscore the critical role of integrating both static and dynamic factors in the development and evaluation of simulation-based approaches, highlighting their combined importance in advancing AV testing methodologies.

% The paper is organized as follows: Section 2 gives a brief idea about the background and concepts we will be using in rest of the paper. Section 3 presents our approach which can be considered new, with respect to existing approaches. Section 4 shows some of the implementation results. Section 5 gives the comparison with the existing approaches. Section 6 concludes this paper and draws future work directions.
The remainder of this paper is organized as follows. Section \ref{sec:background} provides the background necessary to understand the context and key concepts we will be using throughout this work. Section \ref{sec:research_approach} presents the proposed research approach, describing the methodology and techniques used in this study. Section \ref{sec:experimental_results} reports the experimental results and evaluates the effectiveness of the proposed approach. Section \ref{sec:discussion} discusses the implications of the findings and provides further insights into the results. Section \ref{sec:threats} outlines potential threats to the validity of this study, while Section \ref{sec:related_work} reviews related work in the area and positions our contribution within the existing literature. Finally, Section \ref{sec:conclusion} concludes the paper and highlights directions for future research.

\section{Background}
\label{sec:background}

Testing autonomous vehicles presents significant complexity, as evaluation requirements vary depending on the specific system aspects under investigation. Testers may focus on distinct components such as obstacle detection capabilities, driving comfort metrics, or ethical decision-making algorithms, among others. Within the software engineering research community, particular attention has been devoted to evaluating AV lane-keeping systems, as evidenced by numerous studies ~\cite{DBLP:conf/sbst/BiagiolaK24, DBLP:conf/icse/BiagiolaKPR23, GambiJRZ22, DBLP:conf/sbst/PanichellaGZR21, DBLP:conf/icse/GambiMF19}. This focus stems from both the critical safety implications of proper lane maintenance and the system's suitability for controlled testing methodologies.

\subsection{Test Case Definition}

\begin{figure}
    \centering
    \includegraphics[width=0.5\linewidth]{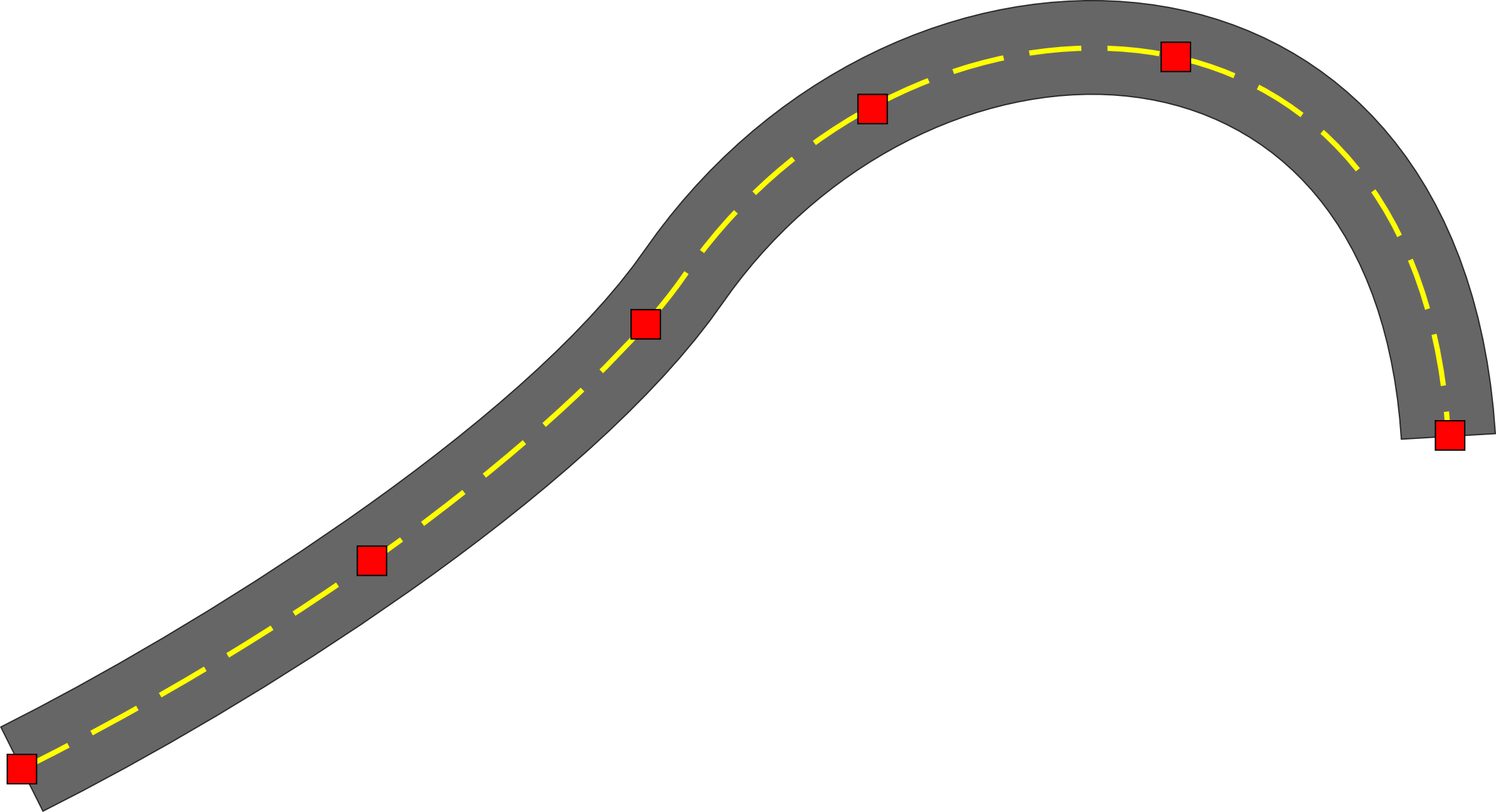}
    \caption{Example of a road generated in the study}
    \label{fig:road}
\end{figure}

In AV testing research, a test case, also referred to as a test scenario, is designed to evaluate the performance of the driving agent responsible for controlling the AV. This usage differs from traditional software testing, where a test case typically refers to a specific set of actions executed to verify a particular feature or functionality, and a test scenario denotes a broader set of test cases aimed at exploring the overall behavior of the system \cite{almog2009test}.

In contrast, AV testing uses both terms interchangeably to describe a dynamic simulation involving the AV and its surrounding environment, both static and dynamic, over time. These scenarios capture the interaction between the vehicle and its environment, including other agents and contextual elements \cite{wang2024survey}.

Testing techniques in AV research differ in how they expose weaknesses in the driving agent, but a common approach is to place the AV in a simulated environment where specific variables are systematically altered to provoke failures. Typical variations include changes in weather conditions \cite{tang2020performance}, the behavior of surrounding road users \cite{crespo2024pafot}, and roadway geometry \cite{frenetic, frenetic_v}.

In this study, we focus on testing the lane-keeping system provided with the \beamng simulator \cite{beamng_tech}. For our purposes, a test case is defined as an ordered sequence of $(X,Y)$ coordinate points in the Cartesian plane. These road points represent the geometric center of the lane and are illustrated as red markers in Figure \ref{fig:road}. The simulation environment constructs the full test case by interpolating and connecting these points, depicted by the yellow dashed line in Figure \ref{fig:road}, to generate the road. It further creates the entire road geometry by calculating and drawing the lanes located in both sides of the interpolated road points.

Our experimental framework builds on the SensoDat dataset \cite{DBLP:journals/corr/abs-2401-09808}, which contains executed test cases specifically designed to evaluate lane-keeping performance in simulated driving environments. Further details of this dataset are provided in Section \ref{sec:dataset}.
% 
% \begin{figure}
%     \centering
%     \includegraphics[width=0.5\linewidth]{road.png}
%     \caption{Example of a road generated in the study}
%     \label{fig:road}
% \end{figure}

\subsection{Performance Metrics of Test Cases}

An essential concept in autonomous-vehicle (AV) testing is the definition of an effective test case. The goal of AV testing is to uncover rare situations in which the vehicle either takes an unsafe action (\eg accelerating at a stop sign) or fails to act appropriately (\eg not braking behind a slower vehicle). Clearly defining what makes a test case effective is therefore critical. Common performance metrics include Time-to-Collision (TTC): the estimated time before the AV would collide with an obstacle if it maintained its current speed and trajectory; Collision count: the total number of collisions with obstacles or other road users; Out-of-Bounds (OOB) incidents: the number of times the AV departs from its designated lane or roadway; Safety-distance violations: instances where the AV follows another vehicle more closely than the prescribed safe distance.

In this study, we use Out-of-Bounds (OOB) incidents as the primary criterion for determining whether a test case is effective or ineffective. Recent research on simulation-based test case generation for AVs typically relies on an oracle to determine whether the system under test behaves safely, using predefined safety metrics \cite{DBLP:journals/ese/BirchlerKBGP23, GambiJRZ22, DBLP:conf/sbst/PanichellaGZR21}. Among these, the OOB metric is one of the most commonly adopted. We selected this metric due to its widespread adoption by both researchers and practitioners, as evidenced in several recent studies \cite{birchler2021automated, DBLP:conf/icst/Jahangirova0T21}. Our evaluation relies on a publicly available dataset that applies the same operational definition (Section \ref{sec:dataset}), ensuring consistency with established benchmarking practices.

%\subsection{Article Terminology}
\subsection{Static Road Features}

Static features define the invariant environmental conditions of simulation-based test cases \cite{DBLP:journals/ese/BirchlerKBGP23}, including:

\begin{itemize}
    \item Infrastructure characteristics: Road geometry, traffic light placement, and sidewalk presence
    \item Environmental elements: weather, nature elements, time of day; provided that they do not change during the test case
    \item Traffic composition: \textit{Number} of vehicles and pedestrians,  static obstacles
\end{itemize}

Our work specifically focuses on Static Road Features - a subset of static features derived exclusively from the test case's road point coordinates. These features, which primarily characterize road shape (e.g., curvature, segment lengths), are predetermined before test execution and align with established metrics in prior research \cite{SDCScissor, DBLP:journals/ese/BirchlerKBGP23}.

\subsection{Dynamic AV Features}
Our approach additionally incorporates Dynamic AV Features – temporal and behavioral data that become available only during or after test case execution. These features capture the spatio-temporal characteristics of both the autonomous vehicle and other road users \cite{neelofar2024identifying}, including:

\begin{itemize}
    \item Kinematic parameters: Position, velocity, and acceleration values for all agents
    \item System state data: Internal operations, sensor readings, and actuator values
    \item Environmental interactions: Real-time responses to dynamic conditions
\end{itemize}

As demonstrated in prior research \cite{zhong2021survey}, the interplay between these dynamic factors (e.g., weather effects, pedestrian movements) and static infrastructure (e.g., road geometry, traffic controls) is critical for detecting safety-critical scenarios. This synergy underscores the importance of monitoring both feature categories during AV testing.

\section{ Research Approach}
\label{sec:research_approach}

This section describes the proposed approach, including the experimental settings, dataset, and parameters used, to answer our research questions.
Specifically, this section offers insights into the dataset used in this study and explains the procedures and methods we apply to implement the ISA methodology as well as the description of ML techniques used to compare the effectiveness of features for predicting the outcome of test cases.

\begin{figure}
    \centering
    \includegraphics[width=\linewidth]{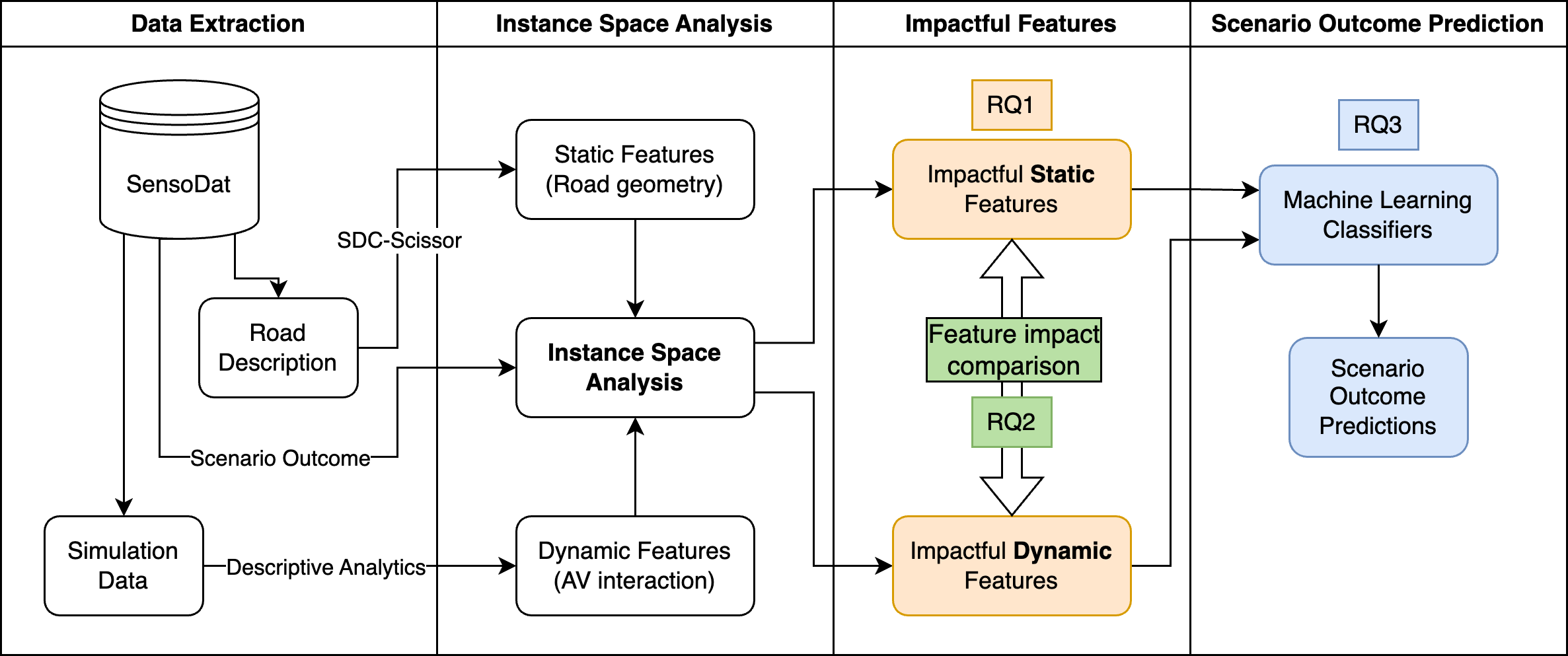}
    \caption{Research Approach Overview}
    \label{fig:approach_overview}
\end{figure}

Figure \ref{fig:approach_overview} illustrates our four-phase research approach to addressing the research questions. The Data Extraction phase focuses on building the dataset necessary for analysis, as outlined in Section \ref{sec:dataset}. In the Instance Space Analysis phase, we construct an instance space using static and dynamic features extracted from simulation data, with test case outcomes defined based on data from \sensodat. The ISA methodology is described in Section \ref{sec:isa}. ISA identifies the most \textit{impactful features}, both static and dynamic, to address RQ1. To answer RQ2, we investigate the interactions and influence between feature types, specifically examining how static features impact AV behavior in test cases, as described by dynamic features. Lastly, the selected impactful features are used to train multiple machine learning classifiers to predict test case outcomes. Details of the machine learning classifiers are presented in Section \ref{sub:ml-classifiers}.

\subsection{Data Extraction}
\label{sec:dataset}

The dataset used in this study consists of test cases from \sensodat~\cite{DBLP:journals/corr/abs-2401-09808}, which evaluates the lane-keeping system agent included in the \beamng\cite{beamng_tech} simulator. 
SensoDat represents a large state-of-the-art dataset for AV test cases. The dataset combines both static road features and dynamic AV features within the same test scenarios and is ideal for our investigation. Unlike existing datasets, which typically include either static or dynamic features but not both, SensoDat provides a comprehensive and integrated perspective by capturing the interactions between road conditions and AV behaviors. This dual-feature inclusion makes SensoDat a valuable resource for analyzing the combined impact of static and dynamic factors, setting it apart from other datasets that focus on only one type of features.
This constitutes the test case subset used in the Instance Space Analysis methodology, explained in the following section.

The test cases represent challenging virtual roads designed to push the AV to drive out of or very close to the lane boundaries. This dataset includes over 32,000 test cases executed in the \beamng simulator. The test cases were generated using three different techniques: Frenetic \cite{frenetic}, FreneticV \cite{frenetic_v}, and AmbieGen \cite{ambiegen}.
Frenetic employs a genetic approach with a curvature-based road representation, utilizing curvatures associated with smooth planar curves to model roads. FreneticV also uses a genetic approach with a curvature-based representation, but it additionally assesses the validity of the generated roads. AmbieGen employs a two-objective NSGA-II algorithm to produce test cases that compel the driving agent to go off the road. These techniques are drawn from the SBST Tool Competition \cite{DBLP:conf/sbst/PanichellaGZR21, GambiJRZ22, DBLP:conf/icse/BiagiolaKPR23}. In this competition, each testing approach must generate a road on which an AV Lane Keeping System is evaluated. The objective is to create roads challenging enough to cause the AV to drift out of its lane. Although certain features, such as intersections, sharp zigzags, or roundabouts, could increase the difficulty, the competition imposes constraints: the road must be continuous, non-self-intersecting, and contained within the virtual map. Because our dataset is derived from this competition, these more complex road features are not included in the analysis.
Each test case is represented as a sequence of $X,Y$ coordinate points, which the simulator interpolates to construct a road. The AV then navigates this road, aiming to reach the end. The dynamic features are derived from sensor data and the internal states of the AV.

From a pool of more than 32,000 test cases, we randomly selected $6,122$ to create a balanced dataset with an equal proportion of effective and ineffective cases across all testing techniques. Within this subset, the Frenetic generator contributed $2,462$ cases, FreneticV produced $1,425$, and AmbieGen supplied $2,235$, for a total of 6,122 test cases. Because effective test cases are rarer than ineffective ones, their overall count is smaller. To create a balanced dataset, we included all effective cases and then randomly selected an equal number of ineffective scenarios from each technique. This method yields an even split between effective and ineffective cases while maintaining the original contribution of each technique relative to the full dataset. The final counts preserve the same proportional representation as in the \sensodat dataset.

A set of static and dynamic features was computed for each test case in this study. Static features were derived using the SDC-Scissor tool \cite{10298301}, producing 19 features per test case. Dynamic features were extracted from the data generated by the sensors onboard the AV, and further refined through descriptive analytics by calculating the maximum, minimum, average, and standard deviation, resulting in 181 dynamic features per test case. Together, these static and dynamic features form the Feature Space, which is then used in the Instance Space Analysis discussed in the next section. The full list of features is available in the repository detailed in Section \ref{sec:data}.

\subsection{Instance Space Analysis}
\label{sec:isa}

Instance Space Analysis (ISA) \cite{Smith-Miles_Munoz_2021} is a relatively new methodology initially proposed for combinatorial optimization problems. However, it has been extended to numerous other fields, such as automated software repair~\cite{aleti2020apr} and automated software testing \cite{Neelofar2023}. ISA maps test instances (\ie test cases), defined by their features, from a $n$-dimensional feature space onto a $2D$ instance space (IS). The projections are created in a way that clearly distinguishes between effective and ineffective test cases, highlighting the influence of each feature on the test outcomes. In the context of testing, ISA helps to understand how the input features of test cases impact their outcomes (\eg effective or ineffective). By mapping these test cases and their features onto a $2D$ space, known as the instance space, ISA provides visual indicators that make it easier to identify this impact. The instance space offers valuable insights into the distribution of existing test cases, highlighting sparse or unoccupied regions where additional test instances can be created to enhance testing comprehensiveness.

For the generation of the instance space, three spaces are needed, which are shown in Figure \ref{fig:ISA_overview}:

\begin{figure}[t]
    \centering
    \includegraphics[width=0.5\linewidth]{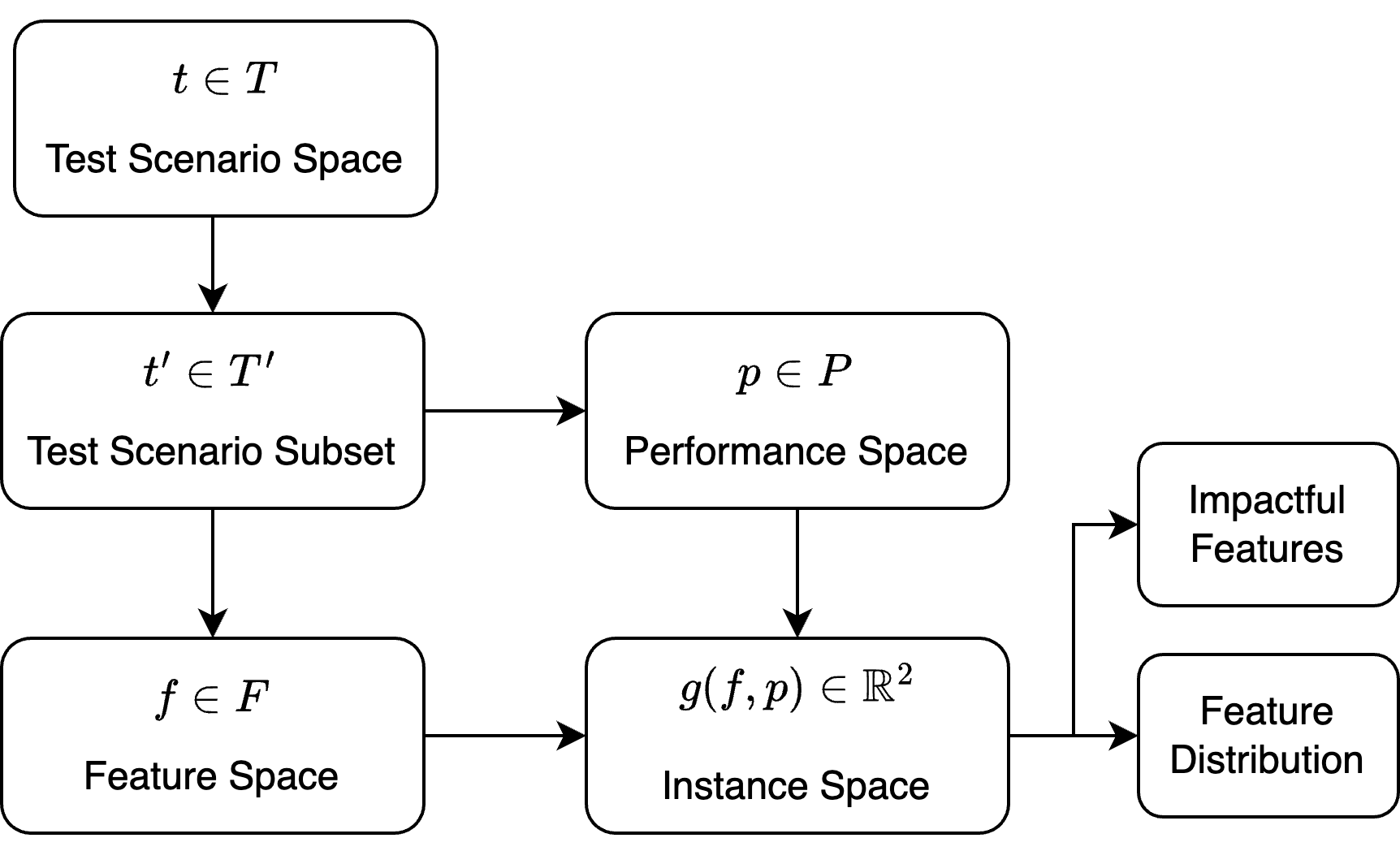}
    \caption{ISA overview}
    \label{fig:ISA_overview}
\end{figure}

\begin{itemize}
    \item Test case Space $T$: This space contains all possible test cases that could be utilized. In our study, this space is formed by all the existing testing techniques focusing on AV testing, including the test cases in the \sensodat dataset (\ie Frenetic, FreneticV, Ambiegen). A subset of test cases denoted as $T'$, is selected from this space to form the test suite used for constructing the instance space. In our case, this space includes the sub-set of $6,122$ test cases selected for analysis.
    \item Feature Space $F$: This space comprises a vector of relevant features that characterize a test case. These features are specific to the domain, and their extraction requires considerable domain expertise \cite{munoz2018instance, munoz2021_instanceregression}. In our study, the Feature Space is created by combining the static features extracted with SDC-Scissor, and the dynamic features extracted by using descriptive analytics. In total, 200 features are included in the construction of this space.
    \item Performance Space $P$: This space represents the performance of the test cases, evaluated using a metric that measures the effectiveness of each test case. It is created by classifying each test case by the occurrence of OOB incidents; a test case where an OOB occurs is labeled as \textit{Effective}, while in an \textit{Ineffective} test case an OOB incident does not occur.
\end{itemize}

An instance space is the $2D$ representation of test cases, defined by the features that have the greatest impact on test case outcomes. Therefore, identifying the features with the most significant impact on the test case outcome is a critical step in ISA. The feature identification and selection process is iterative, using machine learning techniques to uncover key features that clearly distinguish effective from ineffective test cases. An effective test case reveals a bug or incorrect behavior in the system under test. The process starts by calculating the absolute Pearson correlation between each feature and the performance of each algorithm {(\ie Frenetic, FreneticV, Ambiegen)}. It then selects the most correlated feature for each algorithm, along with any other feature that has at least a moderate correlation (\ie above 0.3 \cite{hinkle2003applied}) with at least one algorithm.

Next, a cluster of features with similarities to each other is identified. To do this, a \textit{k}-means clustering algorithm is used to group similar features into clusters. We use a dissimilarity measure of $1-|\rho_{i,j}|$, where $\rho_{i,j}$ is the correlation between two features. We use \textit{k}-means clustering for this process due to its simplicity as an unsupervised machine-learning algorithm. Previous studies \cite{hauer2020clustering} have demonstrated that ISA performs well with this technique. The optimal number of clusters for k-means is determined through silhouette analysis~\cite{aranganayagi2007clustering}; {this analysis quantifies how well a feature fits within its assigned cluster relative to its separation from other clusters. } After the clusters have been formed, one feature is chosen from each cluster to create a feature set. If the previous step resulted in $n$ clusters, the feature set will contain $n$ features. Each $n$-dimensional feature set is then projected into a \textit{temporary} $2D$ space using Principal Component Analysis (PCA)~\cite{abdi2010principal}, providing a simplified view that preserves the most informative variance in the data. This procedure is repeated for all possible combinations of features across all clusters. The coordinates from these \textit{temporary} $2D$ spaces are used as input for a series of Random Forest (RF) models, which identify the feature combinations that minimize predictive error when forecasting test case outcomes.  The resulting subset of features is then utilized to create the instance space. 

With the most effective features identified, we now project the $n$-dimensional feature space into a $2$D coordinate system to clarify the relationship between the features of the test cases and their outcomes. An ideal projection creates a linear trend when examining feature values against test case outcomes, with low values at one end of the line and high values at the other. Additionally, instances that are neighbors in the high-dimensional feature space should remain neighbors in the $2D$ instance space, preserving topological relationships. To achieve this, we employ an optimization method called Projecting Instances with Linearly Observable Trends (PILOT)~\cite{munoz2018instance}. PILOT aims to fit a linear model for each of the impactful features identified in the previous step and test outcome based on location of the instance in the $2D$ plane. Mathematically, this involves solving the following optimization problem:

\begin{eqnarray}
    \label{eq:pilot}
    \min && \left\| \Tilde{\mathbf{F}} - \mathbf{B}_{r}\mathbf{Z} \right\|^{2}_{F} + \left\| \mathbf{Y}^\top - \mathbf{C}_{r}^\top\mathbf{Z} \right\|^{2}_{F} 
    \label{eq:optimisation}
    \\
    \text{s.t.}				&& \mathbf{Z} = \mathbf{A}_{r}\Tilde{\mathbf{F}} \label{eq:score}
    \nonumber
\end{eqnarray}
{The goal of Equation~\ref{eq:pilot} is to minimize the sum of two Frobenius norm-based terms \(\left\| \Tilde{\mathbf{F}} - \mathbf{B}_{r}\mathbf{Z} \right\|^{2}_{F}\) and \(\left\| \mathbf{Y}^\top - \mathbf{C}_{r}^\top\mathbf{Z} \right\|^{2}_{F}\). \(\left\| \Tilde{\mathbf{F}} - \mathbf{B}_{r}\mathbf{Z} \right\|^{2}_{F}\) measures the difference between the original feature matrix \(\Tilde{\mathbf{F}}\) and its estimation \(\mathbf{B}_{r}\mathbf{Z}\). Here $\Tilde{\mathbf{F}} \in \mathbb{R}^{n\times i}$ is a matrix containing the $n$ features of $i$ test scenarios, $\mathbf{Z} \in \mathbb{R}^{2\times i}$ is the matrix containing $z_1$ and $z_2$ coordinate values of $i$ scenarios in the $2D$ space, while $\mathbf{B}_{r} \in \mathbb{R}^{n\times2}$ maps $\mathbf{Z}$ back to the feature space. $\left\| \mathbf{Y}^\top - \mathbf{C}_{r}^\top\mathbf{Z} \right\|^{2}_{F}$ measures the difference between the actual scenario outcomes \(\mathbf{Y}\) and their estimation \(\mathbf{C}_{r}^\top\mathbf{Z}\). Here $\mathbf{Y} \in \mathbb{R}^{i}$ represents a column vector containing the scenario outcomes, while $\mathbf{C}_{r} \in \mathbb{R}^{2}$ maps the $2D$ coordinates $\mathbf{Z}$ to the technique's performance estimation. The constraint $\mathbf{Z} = \mathbf{A}_{r}\Tilde{\mathbf{F}}$
 enforces that the $2D$ coordinates \(\mathbf{Z}\) are obtained by projecting the feature matrix \(\Tilde{\mathbf{F}}\) into a $2D$ space using the matrix $\mathbf{A}_{r} \in \mathbb{R}^{2 \times n}$.}

{In summary, Equation \ref{eq:pilot} seeks the optimal $2D$ representation of the feature matrix \(\Tilde{\mathbf{F}}\), which includes the most impactful features selected in the previous step, by jointly minimizing two sources of error: the reconstruction error of the features, expressed as \(\mathbf{B}_{r}\mathbf{Z}\), and the prediction error of the scenario outcomes, expressed as \(\mathbf{C}_{r}^\top\mathbf{Z}\). In essence, it tries to maintain the structure of the feature space while also making sure that the projected $2D$ coordinates can provide good predictions of the outcomes. Full mathematical proof and additional technical details of the PILOT method can be found in the work of Mu\~{n}oz et al. \cite{munoz_isa_2017, munoz2018instance}.}

\subsubsection{ISA in AV Testing}
\label{sub:isa_av_testing}
In the context of creating the instance space for testing AVs, ISA starts by compiling a diverse set of test cases. These test cases should vary widely in both feature values and execution outcomes (\eg effective and ineffective). The outcome of each test is determined based on a performance metric. In AV testing, common performance metrics include collision probability \cite{lu2022learning}, collision events (\ie collision or no collision) \cite{av-fuzzer}, out-of-bounds episode \cite{birchler2021automated}, and Time to Collision (TTC) \cite{minderhoud2001extended}. In this study we utilize the number of OOB episodes as our performance metric. 

Next, meaningful features are extracted from the test cases, to ensure that they \textit{a)} accurately describe the similarities and differences between test cases; \textit{b)} effectively explain the outcome of the test cases {(\ie they are correlated to the outcome of test cases)}; and \textit{c)} can be computed within a reasonable timeframe. Finally, the instance space is constructed, representing the test cases in a $2D$ space, focusing on features with the biggest impact on the outcome of test cases, as described in Section \ref{sec:isa}. Once the instance space is created, it serves as a powerful tool for evaluating the quality of the test suite that generated it. Moreover, the features identified within the instance space are used to train machine learning models, which can then predict the outcomes of untested test cases without the need for simulations. Details of the machine learning models are presented in Section \ref{sub:ml-classifiers}.

\subsection{Impactful Features}

ISA identifies the features most strongly associated with test-case outcomes. In our study, these features fall into two categories: \textit{static} (\eg road distance, road angles, road curvature) and \textit{dynamic} (\eg steering, control-system activity, vehicle elevation). The features selected for constructing the instance space are chosen based on (a) how well they explain the observed test outcomes and (b) how effectively they separate effective from ineffective test cases in the $2D$ projection. By analyzing both static and dynamic features, we address RQ1.

The interactions among different feature types, and their combined influence on the behavior of the AV and in the outcome of a test case have not been systematically investigated before. ISA addresses this gap by producing a $2D$ representation of test cases in the instance space, while simultaneously generating distributions of the selected feature values that reveal clear linear patterns. This representation, which combines spatial location and feature distributions, enables detailed exploratory data analysis. These visualizations allow us to examine how individual features impact one another and how their interactions shape the likelihood that a test case will be effective or ineffective. The insights drawn from this analysis form the basis for answering RQ2.

\subsection{Scenario Outcome Prediction}
\label{sub:ml-classifiers}

In this study, the static and dynamic features selected by ISA serve as the input data for training machine learning classifiers. These features, specifically chosen by their ability to capture the critical aspects of road geometry and vehicle dynamics, provide a comprehensive dataset for model learning. Our goal is to determine whether the selected features can accurately predict test case outcomes before simulation and compare how effectively each type of feature contributes to this predictive capability. To assess the predictive capability of static and dynamic features, we train five machine learning (ML) models for each type of feature using the features identified as impactful, static and dynamic features. Similarly, we trained five additional models using the whole set of impactful features for comparison. These trained models are then used to predict the outcomes of previous test cases without executing them. The ML classifiers used in this study include Random Forest (RF) \cite{breiman2001random}, Decision Tree (DT) \cite{kotsiantis2013decision}, K-Nearest Neighbors (KNN) \cite{taunk2019brief}, Multilayer Perceptron (MLP) \cite{murtagh1991multilayer}, and Naive Bayes (NB) \cite{rish2001empirical}, which have been considered in previous research on AV testing as relevant ML strategies \cite{ali2019detection, SDCScissor, kruber2019unsupervised}. In total 15 ML models were trained: 5 using static features, 5 using dynamic features, and 5 using both types of features.

To accurately assess the predictive capabilities of the trained models, it is crucial to evaluate them on a new set of test instances that were not used during the training phase. This ensures that the models are not simply memorizing the training data, but are capable of generalizing to unseen test cases. {To ensure a fair comparison, we used a fixed, randomly selected subset of test cases for model training and evaluation. $80\%$ of the dataset ($4,897$ test cases) was allocated for training, while the remaining $20\%$ ($1,224$ test cases) was reserved exclusively for testing and never exposed during training. The same training and testing split was applied consistently across all machine learning classifiers.} We use the Python scikit-learn \cite{kramer2016machine} library for the implementation of the machine-learning models. The performance is measured in terms of precision, recall, and f1-score. This analysis is what we use to address RQ3.

While this study specifically focuses on out-of-bound (OOB) episodes as the performance metric, the proposed methodology is broadly applicable and can be easily extended and adapted to other types of performance metrics, such as collisions or safety violations. The flexibility of the integrated approach, which combines static and dynamic feature analysis, makes it adaptable to diverse AV testing scenarios, as demonstrated in prior work on collision prediction and safety distance violations \cite{neelofar2024identifying, cresporodriguez2024isa}. This adaptability underscores the generalizability of the framework to a wider range of AV testing challenges.

\begin{figure*}[]
\centering
    \includegraphics[width=0.5\textwidth]{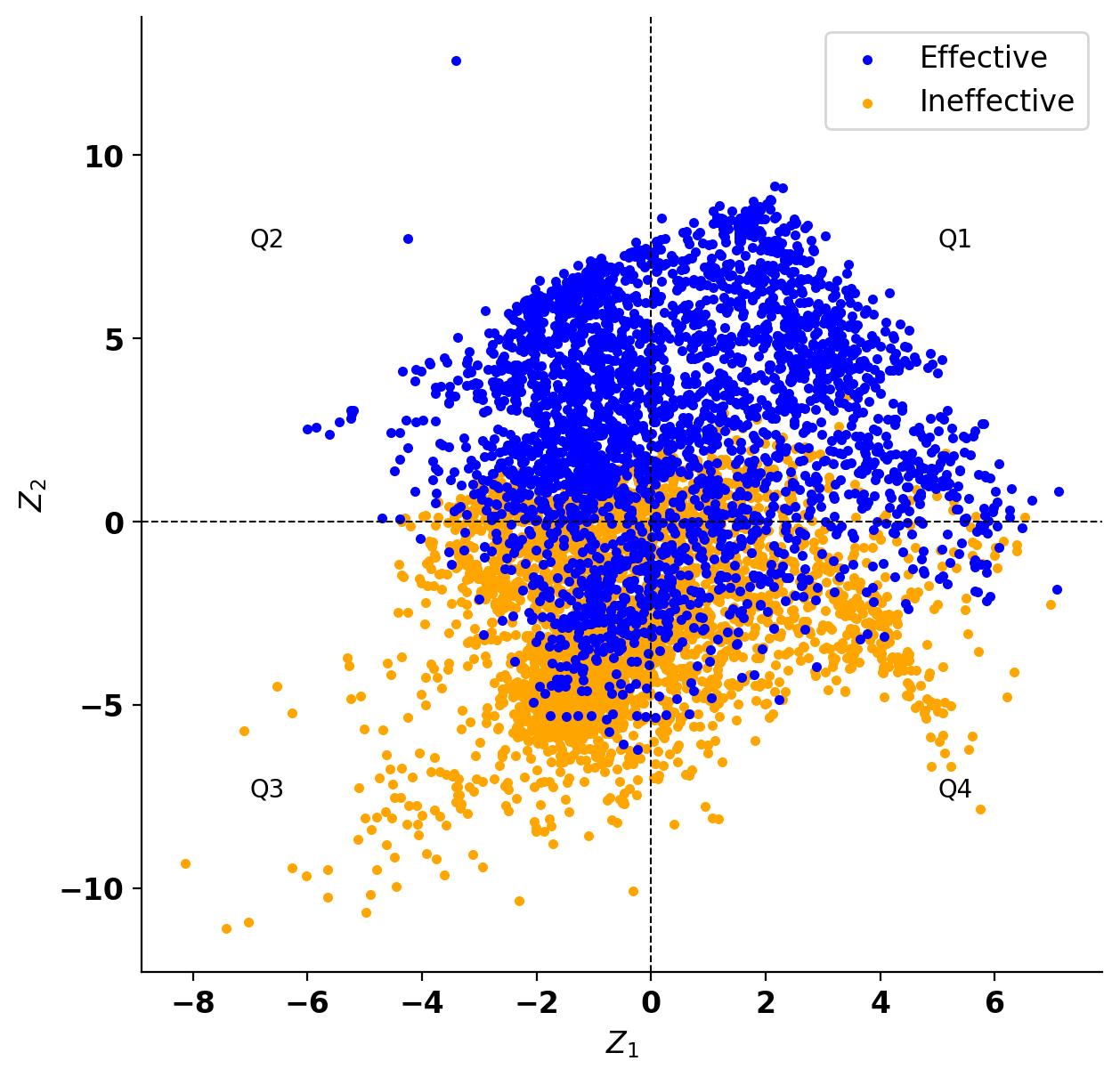}
    \caption{Instance space of test cases}
    \label{fig:instance_space}
\end{figure*}
\section{Results}
\label{sec:experimental_results}

This section provides insights from visualizing feature distributions within the instance space generated by the test suites, addressing our research questions from Section \ref{sec:introduction}. Figure \ref{fig:instance_space} shows this instance space, with test cases classified as effective (OOB incident, in blue) or ineffective (no OOB incident, in orange). Each point represents a test case generated by the dataset's testing techniques, while axes $Z_1$ and $Z_2$ show the principal components. ISA appears to effectively separate effective and ineffective cases, creating distinct clusters with shared characteristics. This instance space will be referenced in the following sections.

\subsection{Key static and dynamic features (RQ1)}
\label{sec:results_rq1}

Of the features forming the Feature Space explained in Section \ref{sec:isa}, ISA selected 16 features after pre-processing, {shown in Table \ref{tab:impactful_features}. Of these features 10 are dynamic features and 6 are static}. The features discarded in this step showed very little or no correlation with the outcome of test cases. The retained features each exhibited a correlation value $\rho \ge 0.3$. This correlation value was selected empirically after experimenting with different values and presents an adequate balance between the number of features selected and the correlation with the outcome of test cases. In the next step, 10 features were kept after clustering to find the features that best describe the outcome of test cases. The clusters generated by the combination of these 10 features showed the lowest predictive error when forecasting test case outcomes, and were selected as the most impactful features of the study. The process of selection and clustering is detailed in Section \ref{sec:isa}. These selected features are associated with the effectiveness of a test case and are presented {as F1 to F10 in} Table \ref{tab:impactful_features}.  Five of these features correspond to dynamic features, and the remaining five are static features.

\begin{table}[!h]
    \centering
    \caption{Most impactful features}
    \small
    \begin{tabular}{l l l p{14em}}\toprule
    \textbf{Feature} & \textbf{Correlation} & \textbf{Type} & \textbf{Description}\\
    \midrule
        F1: steering\_std           & $0.3914$   & Dynamic   & Standard deviation of the steering values \\
        F2: steering\_min           & $-0.4154$   & Dynamic   & Minimum value of steering \\
        F3: steering\_input\_max    & $0.4354$   & Dynamic   & Maximum value of the steering input \\
        F4: altitude\_max           & $0.4729$   & Dynamic   & Maximum altitude experienced during the test case \\
        F5: esc\_std                & $0.3505$   & Dynamic   & Standard deviation of the number of times the Electronic Stability Control (ESC) was activated during the test case\\
        F6: full\_road\_diversity   & $-0.4589$   & Static   & Total area covered by the curves in the road\\
        F7: max\_angle              & $-0.3549$   & Static   & Maximum angle of the road\\
        F8: min\_angle              & $0.3574$   & Static   & Minimum angle of the road\\
        F9: road\_distance          & $-0.7103$   & Static   & Total distance of the road\\
        F10: std\_angle             & $-0.4159$   & Static   & Standard deviation of the angles in the road \\
        F11: num\_l\_turns          & $-0.5891$   & Static   & Number of left turns in the road \\ 
        F12: altitude\_mean         & $0.4733$   & Dynamic   & Average value of altitude experienced during the test case \\
        F13: steering\_input\_std   & $0.4134$   & Dynamic   & Standard deviation of the steering input \\
        F14: esc\_active\_mean      & $0.3581$   & Dynamic   & Average time of ESC being active \\
        F15: esc\_active\_std       & $0.3571$   & Dynamic   & Standard deviation of the time when ESC is active \\
        F16: esc\_mean              & $0.3509$   & Dynamic   & Average value of the number of times the ESC was activated \\
        \bottomrule
    \end{tabular}
    \label{tab:impactful_features}
\end{table}
\normalsize

\subsubsection{Static Features}

 Static features include the characteristics that describe the geometry of a road, such as lane structures, road curves, and intersections. These features remain constant throughout the test case. The most impactful static features selected by ISA include \texttt{full\_road\_diversity} {which describes the cumulative area spanned by all curves in the road}; \texttt{max\_angle}, \texttt{min\_angle}, and \texttt{std\_angle} which describe the maximum, minimum, and standard deviation {of the turn angles of the road}; and \texttt{road\_distance} which describes the total length of the road used to run the test case. The selected features are then projected onto a $2D$ instance space. The projection matrix defined by the linear transformations is shown in Equation \ref{eq:pca_npcs}, highlighting the contribution of each feature to the $z_1$ and $z_2$ axis. \texttt{max\_angle} has the highest contribution to $z_1$, while \texttt{std\_angle} contributed the most to $z_2$.

\subsubsection{Dynamic features}

Dynamic features describe the driving operations of the AV, including its speed, acceleration, steering, and a range of sensors on board of the AV. These features vary throughout the test case as the AV interacts and responds to the environment. The most impactful dynamic features selected by ISA are \texttt{steering\_std}, describing the standard deviation of the steering values effected by the AV; \texttt{steering\_min}, which presents the minimum steering value during the execution; \texttt{steering\_input\_max} which describes the maximum input to the steering performed by the AV; \texttt{altitude\_max} which describes the elevation of the AV relative to the road; and \texttt{esc\_std} which describes the standard deviation of the operations performed by the Electronic Stability Control (ESC) unit within the AV. Similar to static features, dynamic features are projected onto a $2D$ instance space. The projection matrix shown in Equation \ref{eq:pca_npcs} highlights the contribution of each feature to the $z_1$ and $z_2$ axis. \texttt{steering\_min} has the highest contribution to $z_1$, and \texttt{altitude\_max} contributes the most to $z_2$.

\begin{equation}
\label{eq:pca_npcs}
\begin{bmatrix}
	z_1 \\
	z_2
\end{bmatrix}
= 
 \begin{bmatrix}
	{0.667} &   {0.359}   \\
        {-0.696} &   {-0.511} \\
        {0.655} &   {0.401}  \\
        {0.253}  &   {0.549}  \\
        {0.287} &   {0.391}\\
        {0.302}  &   {-0.567} \\
        {0.402}  &   {-0.581}  \\
        {-0.202} &   {0.591}  \\
        {0.213} &   {-1.026} \\
        {0.384} &   {-0.606}  

\end{bmatrix}^{T}
 \begin{bmatrix}
	{\text{ steering\_std }}  \\
	{\text{ steering\_min }} \\
        {\text{ steering\_input\_max }}\\
	{\text{altitude\_max}}\\
	{\text{esc\_std}}\\
	{\text{full\_road\_diversity}}\\
	{\text{max\_angle}}\\
	{\text{min\_angle}}\\
	{\text{road\_distance}}\\
        {\text{std\_angle}}
\end{bmatrix}
\end{equation}

\begin{custombox}
    \textbf{Answer to RQ1:} Static and dynamic features of a test case play a role in generating effective test cases, though not all features contribute equally to their effectiveness. ISA identified a subset of static and dynamic features that most significantly impact the effectiveness of test cases. The \textit{Static Features}, which capture road geometry and characteristics, include \texttt{full\_road\_diversity}, representing the variety of curves along the road; \texttt{max\_angle}, \texttt{min\_angle}, and \texttt{std\_angle}, describing the road angles; and \texttt{road\_distance}, which relates to the road’s length. Conversely, \textit{Dynamic Features} include \texttt{steering\_std}, \texttt{steering\_min}, and \texttt{steering\_input\_max}, which describe the AV’s steering behavior; \texttt{altitude\_max}, indicating the AV’s elevation; and \texttt{esc\_std}, which reflects the operation of the Electronic Stability Control system.
\end{custombox}

\subsection{Impact of road features on AV behavior (RQ2)}
\label{sec:results_rq2}

\begin{figure*}[t]
    \begin{subfigure}[t]{0.3\textwidth}
        \includegraphics[width=\textwidth]{scenario_outcome.png}
        \caption{Test case Outcome}
        \label{fig:isa_dist_sta_outcome}
    \end{subfigure}
    \hfill
    \begin{subfigure}[t]{0.3\textwidth}
        \includegraphics[width=\textwidth]{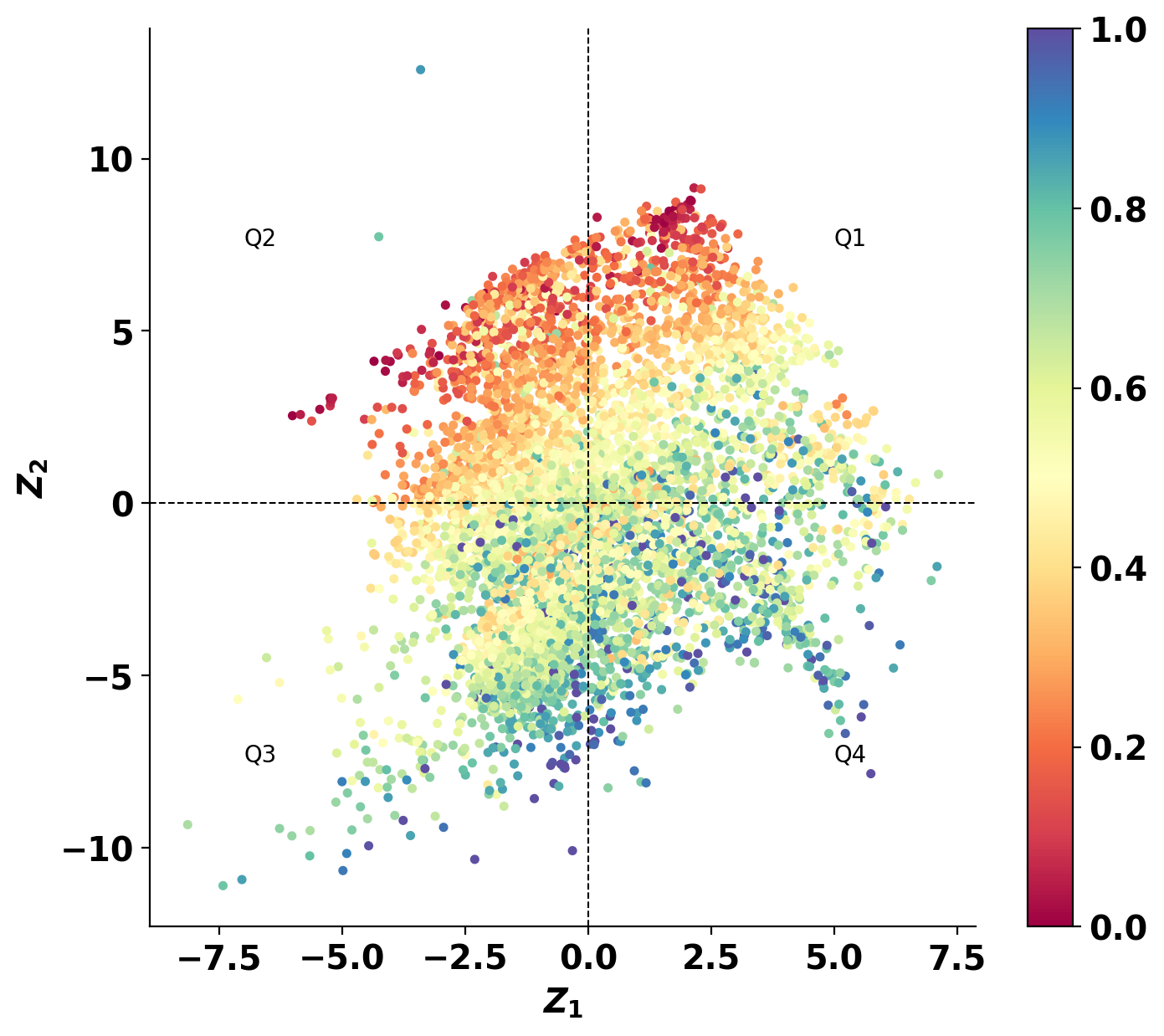}
        \caption{Full Road Diversity}
        \label{fig:isa_dist_sta_road_diversity}
    \end{subfigure}
    \hfill
    \begin{subfigure}[t]{0.3\textwidth}
        \includegraphics[width=\textwidth]{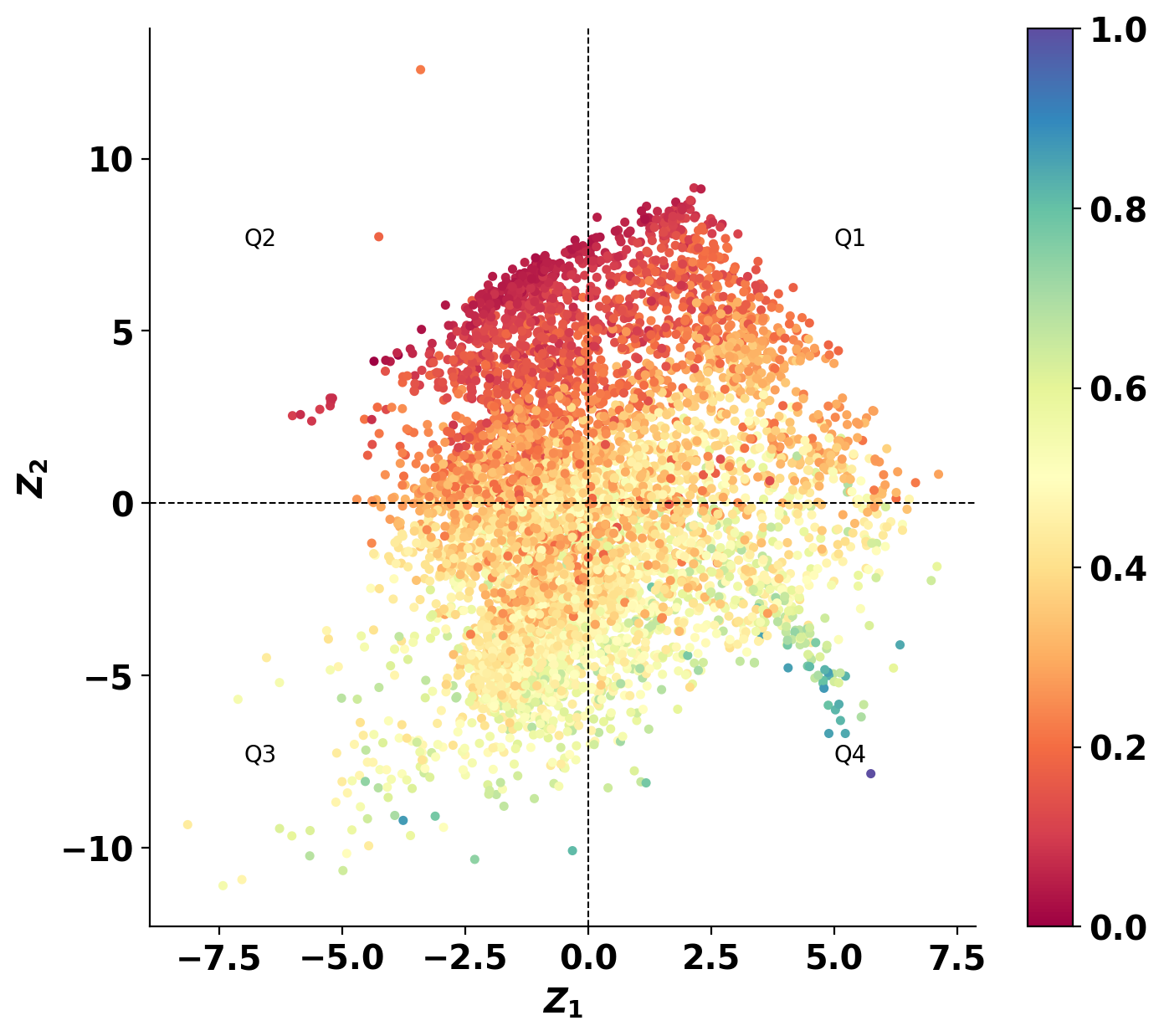}
        \caption{Road Distance}
        \label{fig:isa_dist_sta_road_distance}
    \end{subfigure}
    \begin{subfigure}[t]{0.3\textwidth}
        \includegraphics[width=\textwidth]{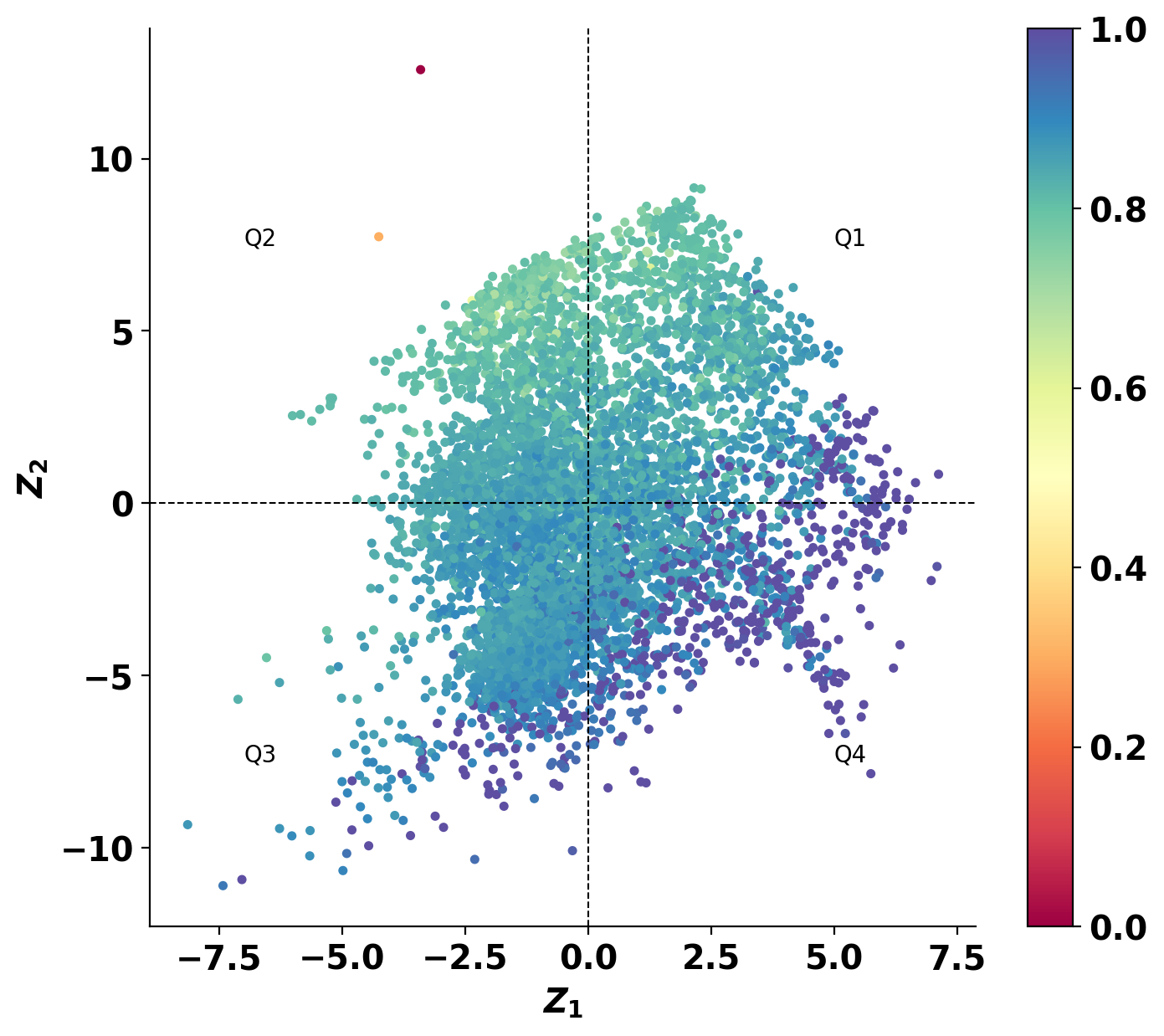}
        \caption{Max Angle}
        \label{fig:isa_dist_sta_max_angle}
    \end{subfigure}
    \hfill
    \begin{subfigure}[t]{0.3\textwidth}
        \includegraphics[width=\textwidth]{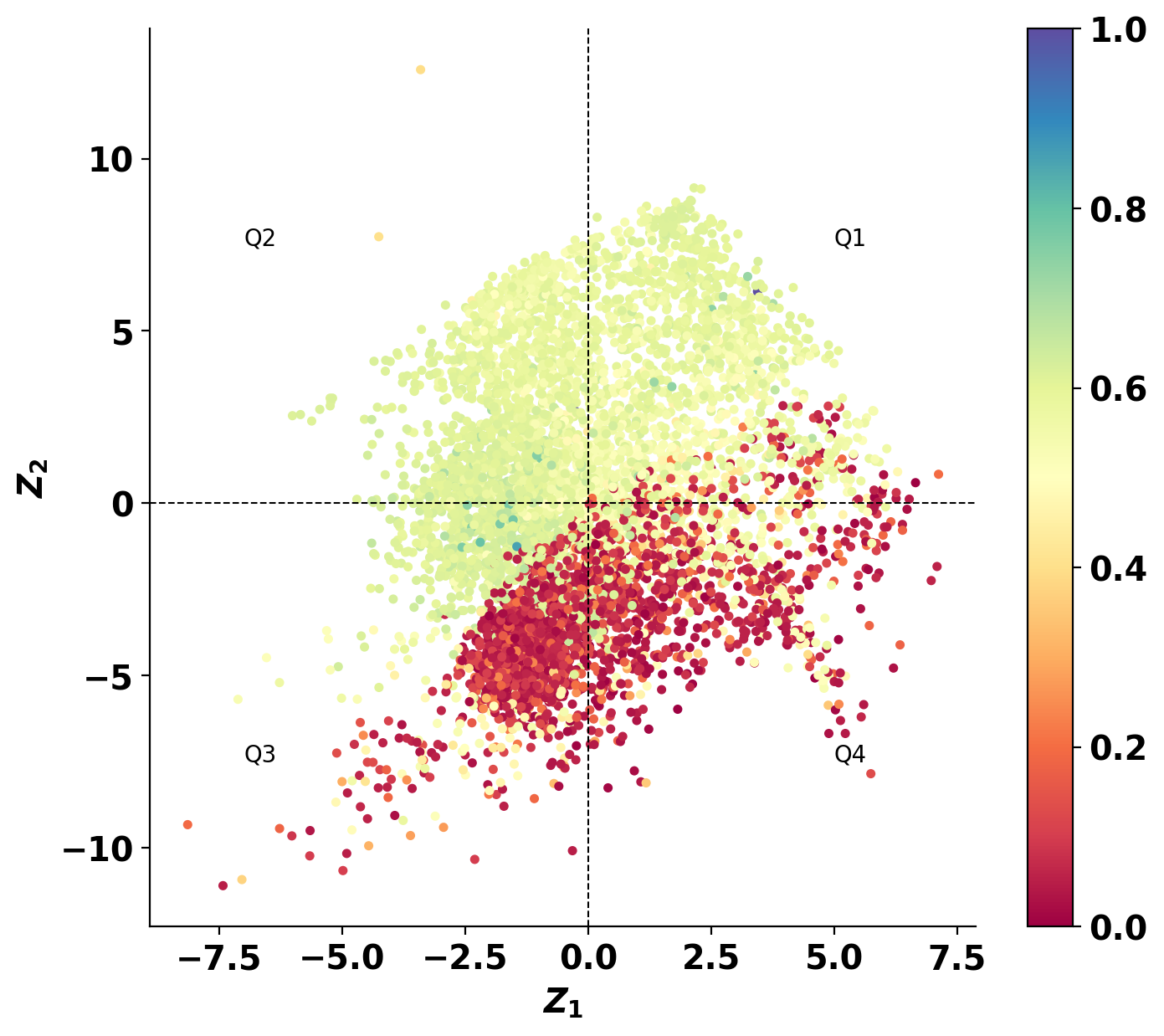}
        \caption{Min Angle}
        \label{fig:isa_dist_sta_min_angle}
    \end{subfigure}
    \hfill
    \begin{subfigure}[t]{0.3\textwidth}
        \includegraphics[width=\textwidth]{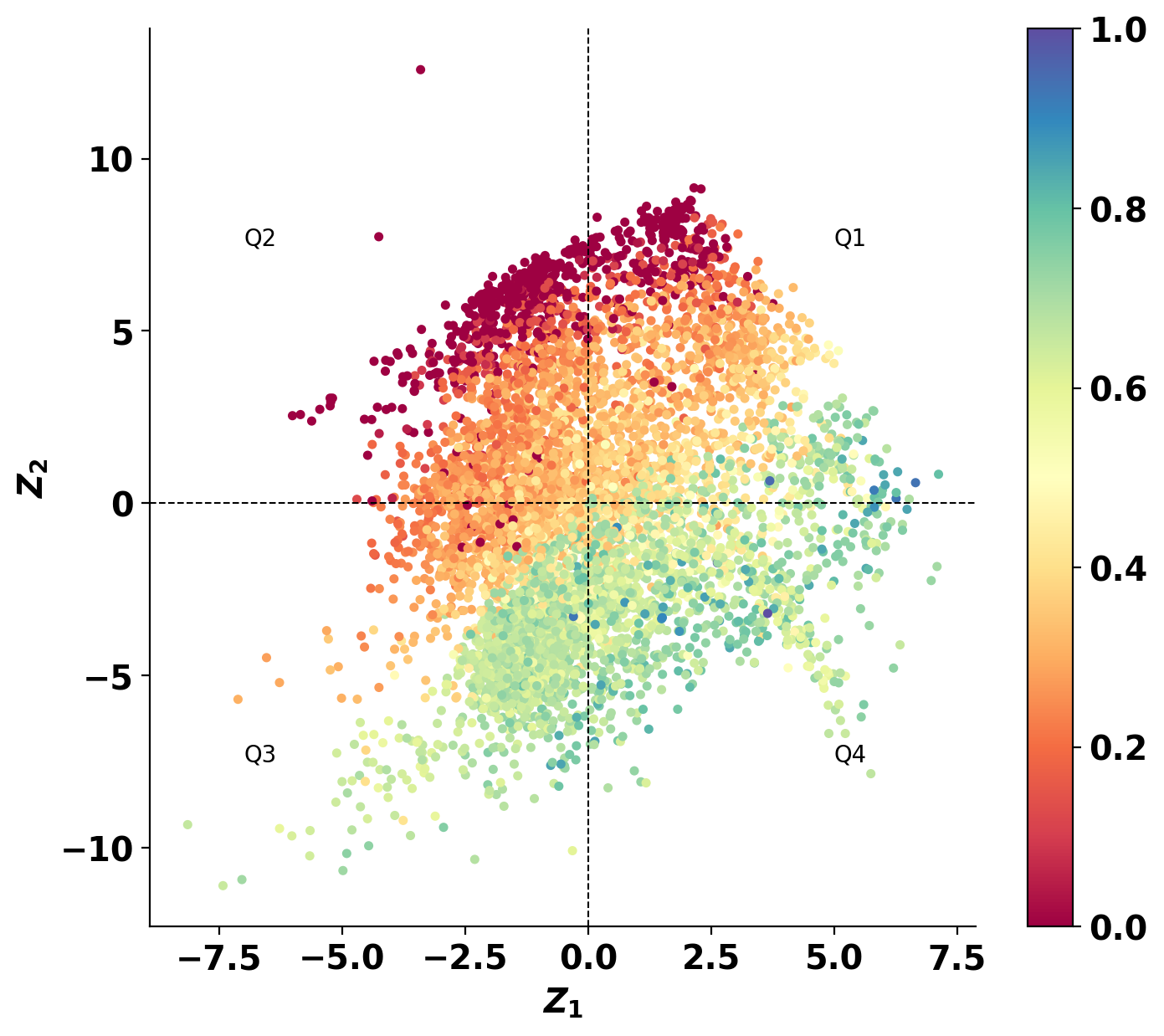}
        \caption{Std Angle}
        \label{fig:isa_dist_sta_angle_std}
    \end{subfigure}
        \caption{Distribution of values of static features}
    \label{fig:isa_distribution_static}
\end{figure*}

\begin{figure*}[t]
    \begin{subfigure}[t]{0.3\textwidth}
        \includegraphics[width=\textwidth]{scenario_outcome.png}
        \caption{Test case Outcome}
        \label{fig:isa_dist_dyn_outcome}
    \end{subfigure}
    \hfill
    \begin{subfigure}[t]{0.3\textwidth}
        \includegraphics[width=\textwidth]{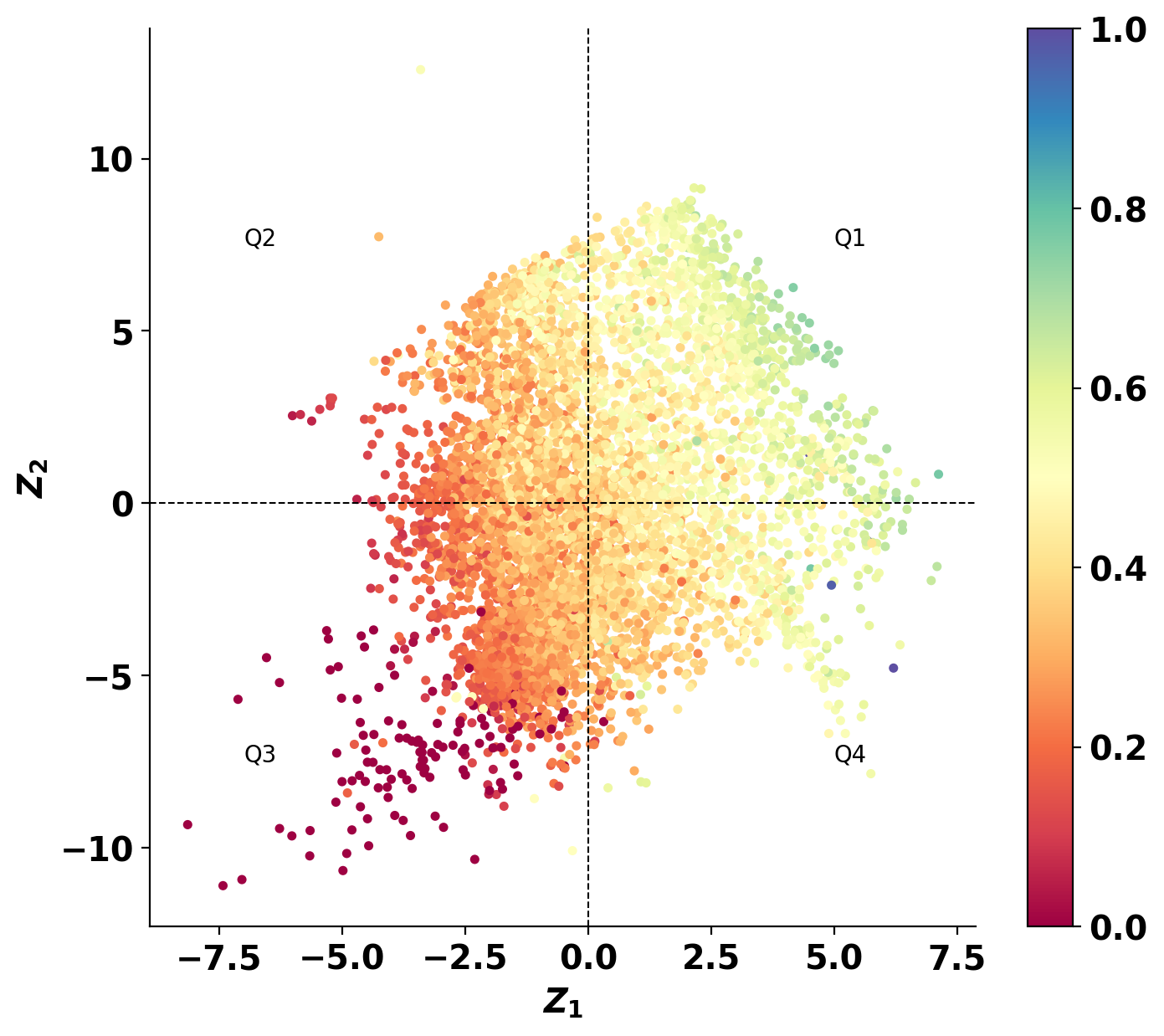}
        \caption{Steering Std}
        \label{fig:isa_dist_dyn_steering_std}
    \end{subfigure}
    \hfill
    % \label{fig:enter-label}
    \begin{subfigure}[t]{0.3\textwidth}
        \includegraphics[width=\textwidth]{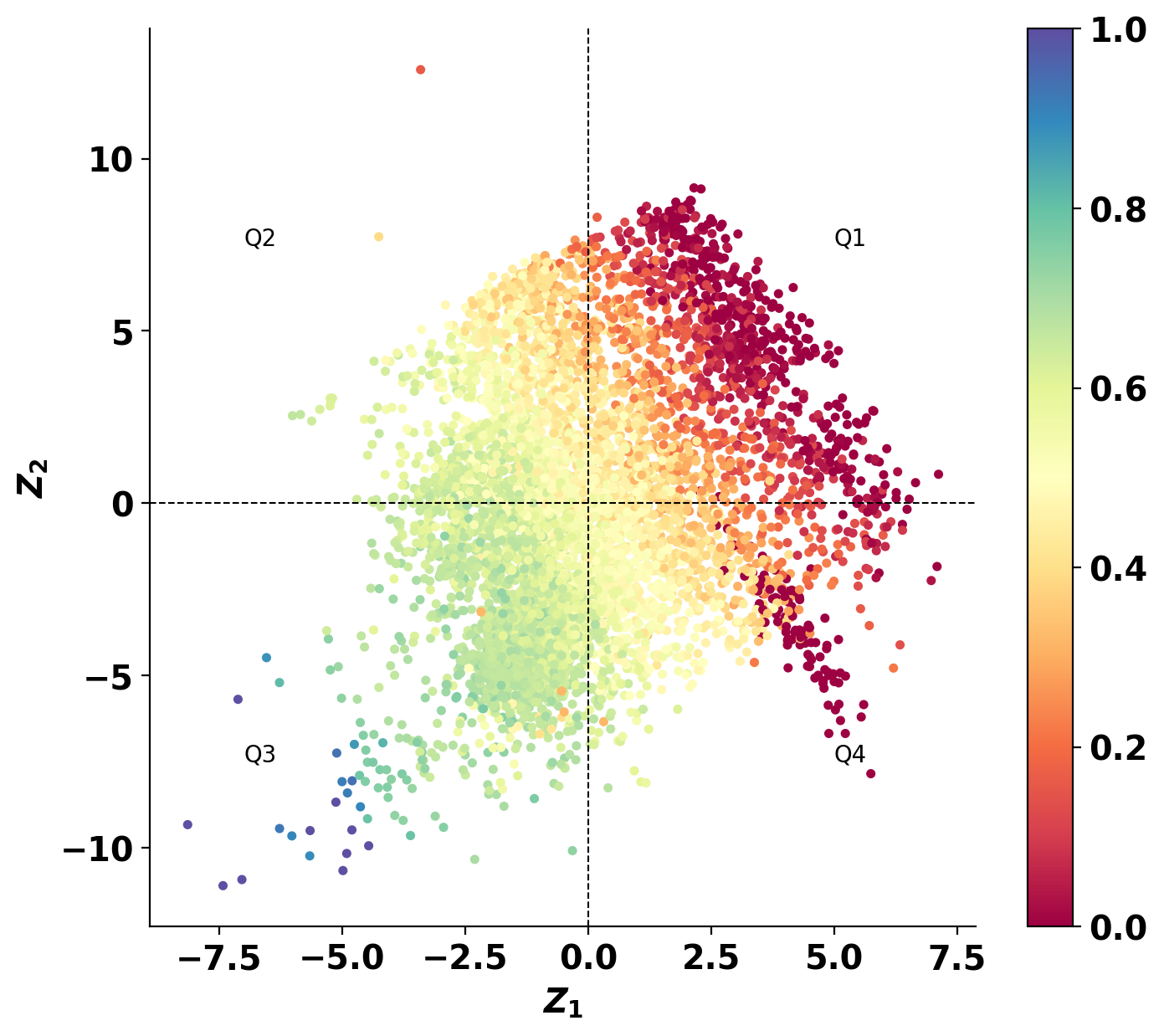}
        \caption{Steering Min}
        \label{fig:isa_dist_dyn_steering_min}
    \end{subfigure}

    \begin{subfigure}[t]{0.3\textwidth}
        \includegraphics[width=\textwidth]{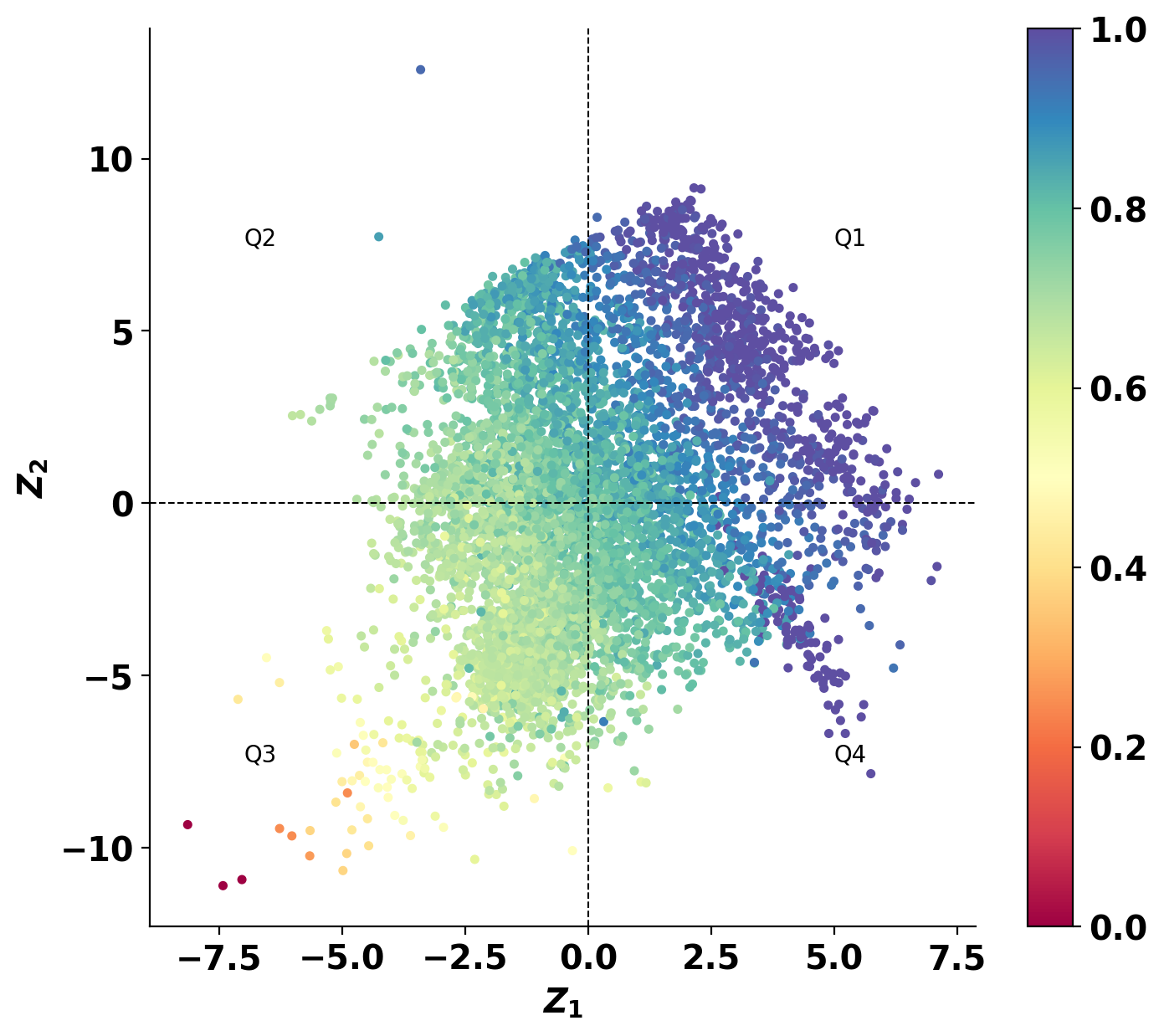}
         \caption{Steering Input Max}
         \label{fig:isa_dist_dyn_steering_max}
    \end{subfigure}
    \hfill
    \begin{subfigure}[t]{0.3\textwidth}
        \includegraphics[width=\textwidth]{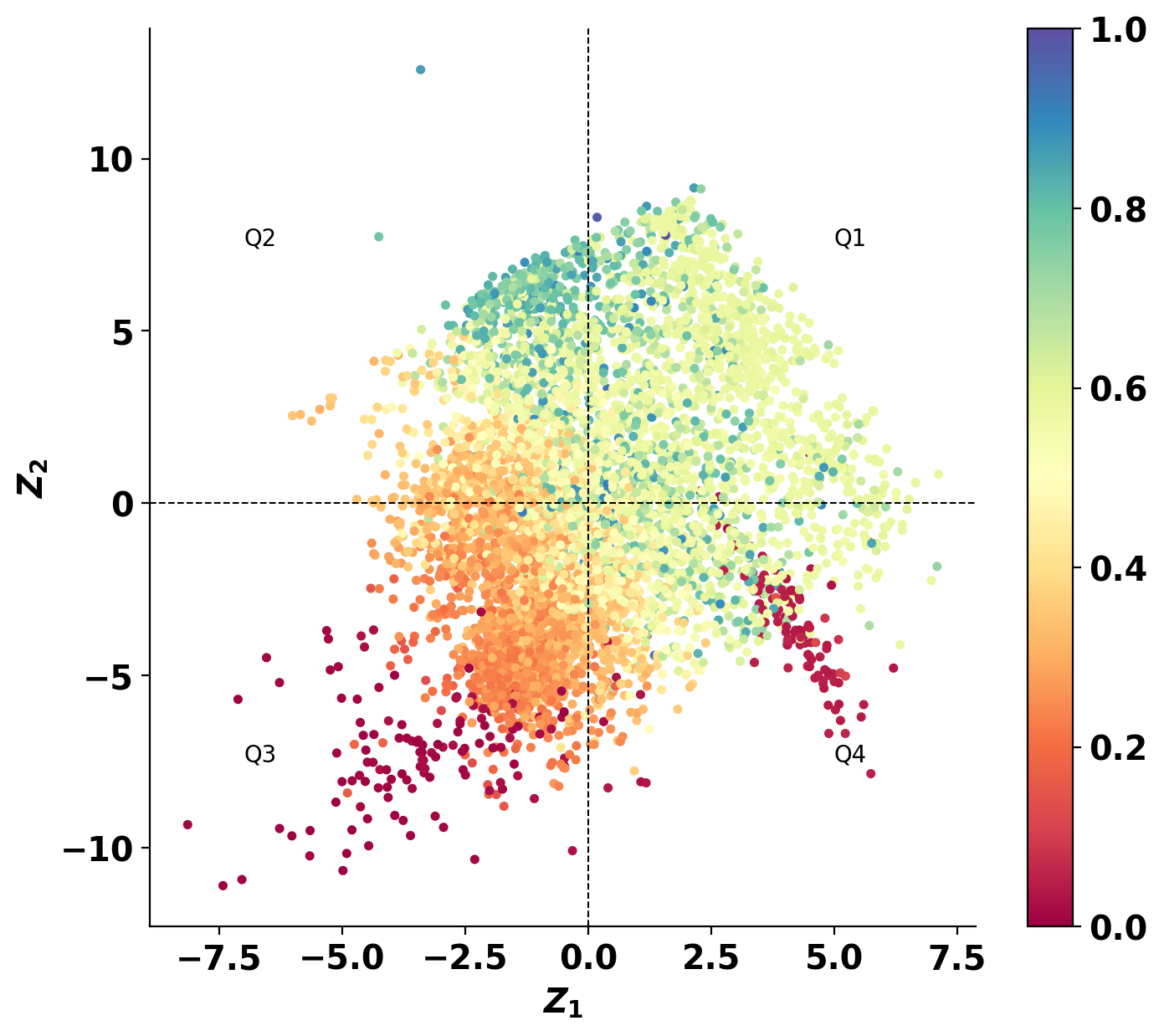}
        \caption{Altitude Max}
        \label{fig:isa_dist_dyn_altitude_max}
    \end{subfigure}
    \hfill
    \begin{subfigure}[t]{0.3\textwidth}
        \includegraphics[width=\textwidth]{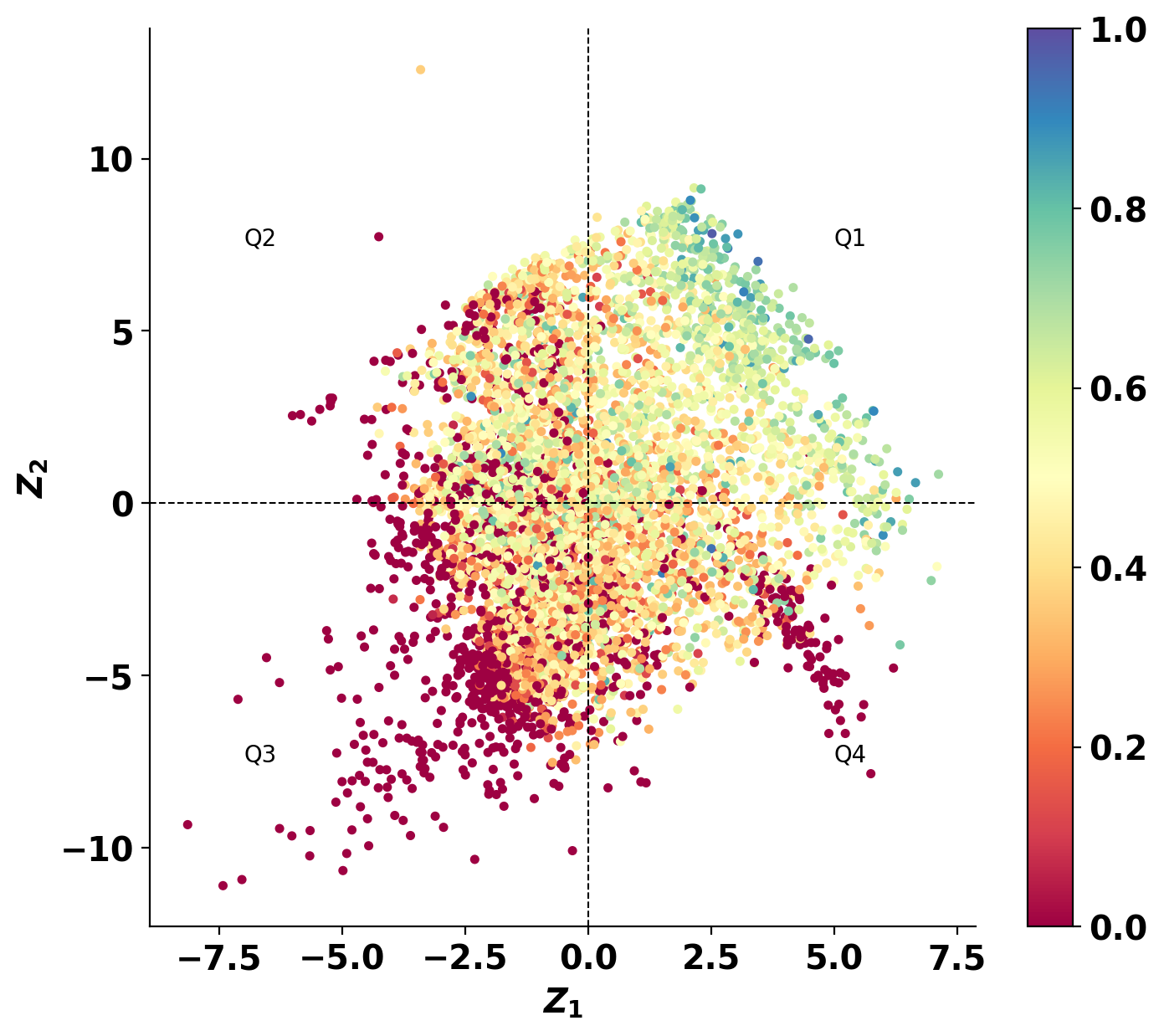}
        \caption{ESC Std}
        \label{fig:isa_dist_dyn_esc_std}
    \end{subfigure}  
    \caption{Distribution of values of dynamic features}
    \label{fig:isa_distribution_dynamic}
\end{figure*}

Static features play a critical role in shaping AV driving behavior, directly influencing its dynamic features. For instance, sharper curves or narrow lanes may lead to more frequent and pronounced steering adjustments, while elevation changes could affect its speed and stability. By understanding how these static elements interact with dynamic operations, we can gain deeper insights into the factors that influence the overall performance of AVs and safety during testing. In addition to selecting the most impactful static and dynamic features for effective test cases, ISA produces a 2D visualization of feature value distributions across the instance space. These distributions facilitate exploratory data analysis by revealing clusters of effective test cases and the values of features associated with them. Figure \ref{fig:isa_distribution_dynamic} presents the distribution of values of dynamic features, and Figure \ref{fig:isa_distribution_static} shows the distribution of values of static features. Figures \ref{fig:isa_dist_sta_outcome} and \ref{fig:isa_dist_dyn_outcome} present the location of effective and ineffective test cases, denoted as blue and orange dots.

These figures display the normalized value of each feature for every test case, positioned according to its coordinates in the 2D instance space. Feature values are mapped so that gradual changes, such as smooth gradients or distinct clusters, are visually apparent, allowing potential linear trends to emerge. In many cases, lower feature values appear at one end of the instance space while higher values concentrate toward the opposite end.

By comparing the spatial distribution of a feature’s values with the location and outcome of each test case, we can identify how specific feature ranges influence whether a scenario is effective or ineffective. It is important to note that all values are normalized to the $[0.0,1.0]$ interval and therefore represent relative magnitudes rather than physical units. For example, a maximum‐angle value of $0.7$ does not indicate a road curvature of 0.7 degrees; it simply means that this test case’s maximum angle lies at $70\%$ of the highest value observed across the entire dataset.

 Effective test cases are primarily concentrated in quadrants 1 and 2 of the instance space, as seen in Figures \ref{fig:isa_dist_dyn_outcome} and \ref{fig:isa_dist_sta_outcome}. The most impactful dynamic features involve AV steering behavior (\texttt{steering\_std}, \texttt{steering\_min}, \texttt{steering\_input\_max}), altitude (\texttt{altitude\_max}), and ESC performance (\texttt{esc\_std}). These features are shaped by interactions with road-related static features, including curvature angles (\texttt{max\_angle}, \texttt{min\_angle}, \texttt{std\_angle}), curve diversity (\texttt{full\_road\_diversity}), and road length (\texttt{road\_distance}).

% \restoregeometry

Effective test cases are associated with mid-to-high values of \texttt{max\_angle} ($0.6$ to $0.8$), indicating more pronounced curves, while lower values suggest gentler curves. For \texttt{min\_angle}, effective test cases typically fall around medium values ($\approx0.5$); a value of $0$ indicates a straight road segment, and values near $1$ suggest continuous curves. These angles are normalized, reflecting their distribution rather than actual degrees. 

The \texttt{std\_angle} feature captures the variability of road angles. Low \texttt{std\_angle} values indicate that the road’s angles are relatively uniform, which occurs when the route is largely straight or follows a long, continuous curve. In contrast, higher \texttt{std\_angle} values reflect a mix of bends and sharper turns, signaling a more complex road geometry.

The \texttt{full\_road\_diversity} feature measures the overall variety of straight and curved segments along the route. Effective test cases typically exhibit low-to-moderate diversity, suggesting roads that combine straight stretches with occasional curves. High values of this feature correspond to roads dominated by large, sweeping curves, whereas low values indicate a nearly continuous straight segment.

Finally, the feature \texttt{road\_distance} does not directly influence AV operation but plays a role in defining effective test cases, as shorter roads are more commonly associated with effective cases. However, it is essential to note that \texttt{road\_distance} alone does not determine test case effectiveness; a short test road does not necessarily imply an effective test case on its own.

The influence of these features on AV performance is closely linked to its steering operations. Three key metrics describe the AV steering behavior during test cases: \texttt{steering\_std} (the standard deviation of steering throughout the test case), \texttt{steering\_min} (the minimum steering angle), and \texttt{steering\_input\_max} (the maximum input to the steering actuator). Steering adjustments are typically necessary when navigating curves, as the AV must maneuver within its lane to avoid OOB incidents. In contrast, straight roads do not require significant steering input. Effective test cases often correlate with medium-to-high values of \texttt{steering\_std}, indicating substantial steering variability during execution. Lower values of \texttt{steering\_min} suggest that the AV maintained a relatively straight trajectory, while high values of \texttt{steering\_input\_max} indicate that considerable steering adjustments were required.

{The feature \texttt{altitude\_max} measures the highest vertical position (elevation) of the autonomous vehicle (AV) during a test. In test cases that are considered effective, the AV often reaches medium to high elevation values. This indicates that in effective test cases the car takes fast and sharp turns; as a result the sideways (lateral) forces can make it tilt or ``roll'' slightly, and the sensors capture this small change in altitude.} Furthermore, the \texttt{esc\_std} feature measures the frequency of Electronic Stability Control (ESC) activation. Mid-to-high values suggest frequent transitions between on and off states, whereas low values indicate that the ESC remained consistently engaged or was unnecessary in less effective test cases. The ESC is activated when the conditions of driving operations could impact the stability of the AV.

Together, these features offer a clearer understanding of driving dynamics in effective test cases: roads in these test cases consist of long, straight segments followed by sharp curves. This layout causes the AV to drive at higher speeds along extended straight sections, followed by abrupt steering adjustments that often lead to out-of-bounds incidents.

\begin{custombox}
    \textbf{Answer to RQ2:} Static features play a crucial role in influencing dynamic features, thereby shaping the overall performance and behavior of the AV. In our study, road curvature, as defined by static angle features, directly affects the AV steering operations, which are captured by dynamic features such as the minimum, maximum, and standard deviation of steering angles. Additionally, data from the altitude sensor provides further insights into how road characteristics impact the AV's steering behavior. Although they do not directly influence AV operations, static features related to road length also affect the outcomes of test cases. By examining the interplay between static and dynamic features, we can better understand the types of roads and AV behaviors that contribute to OOB incidents during AV testing.
\end{custombox}

\subsection{Regression testing: Predicting test case outcome (RQ3)}
\label{sec:results_rq3}

Table \ref{tab:ml_performance} summarizes the performance of five machine learning models—Random Forest (RF), Decision Tree (DT), K-Nearest Neighbors (KNN), Multilayer Perceptron (MLP), and Naive Bayes (NB)—using three standard evaluation metrics: precision (P), recall (R), and F1 score (F1). Precision captures the proportion of correctly predicted positive instances among all positive predictions, indicating how often the positive predictions of the model are correct. Recall reflects the proportion of actual positive instances that the model successfully identifies, highlighting its ability to detect true positives. The F1 score, the harmonic mean of precision and recall, provides a single balanced measure, particularly valuable when precision and recall must be considered jointly or when they trade off against each other \cite{goutte2005probabilistic}.

Each model was evaluated under three configurations: trained exclusively on dynamic features, exclusively on static features, and on the full feature set combining both. The rows labeled Dynamic, Static, and All correspond to these configurations. Models trained on dynamic features relied on variables capturing motion and behavioral characteristics of the scenario—specifically \texttt{steering\_std}, \texttt{steering\_min}, \texttt{steering\_input\_max}, \texttt{altitude\_max}, and \texttt{esc\_std}. Those trained on static features used variables describing structural and geometric properties of the environment, including \texttt{full\_road\_diversity}, \texttt{max\_angle}, \texttt{min\_angle}, \texttt{std\_angle}, and \texttt{road\_distance}. The combined configuration incorporated the entire set of static and dynamic features to assess whether integrating both types of information yields better predictive performance.

Among these, Random Forest (RF) achieves the highest performance, with a precision of $0.958$, recall of $0.915$, and F1 score of $0.936$, underscoring its effectiveness in this testing context. {It is important to note that, although the RF classifier achieved the highest overall performance, the primary goal of this study was not to crown the best individual model but to evaluate whether combining static and dynamic features improves predictive accuracy compared with using only one feature type. In this respect, models trained on the combined feature set consistently outperformed those trained solely on static or dynamic features, demonstrating a clear advantage in utilizing both feature types together.}

The results show that, for the dataset used in this study, models trained with a combination of static and dynamic features perform significantly better than those using only one feature type, either static or dynamic features. These features can therefore be applied in test case selection and prioritization techniques, enabling the identification of critical test cases based on impactful features without requiring execution in a simulator. Models trained on a single feature type, while performing relatively well, fall short compared to the significantly improved performance observed when using both feature types together. Classifiers trained with dynamic features alone achieve over 70\% in precision, recall, and F1 scores, and those using only static features exceed 80\%. However, models trained with the full set of static and dynamic features consistently demonstrate superior results when compared to their single feature-type counterparts, underlining the critical advantage of using \textit{Static and Dynamic Features}. This strong performance difference emphasizes the importance of using both feature types together, solidifying confidence in the relevance and suitability of the selected features for accurately identifying critical test cases.

\begin{custombox}
    \textbf{Answer to RQ3:} Machine learning classifiers trained with both static and dynamic features significantly outperform those that rely on only one type, achieving much higher \textbf{Precision}, \textbf{Recall}, and \textbf{F1-scores}. The Random Forest (RF) model achieves the highest performance, with a precision of $0.958$, recall of $0.915$, and F1 score of $0.936$. These findings reinforce the necessity of integrating both types of features when training ML models, as doing so greatly improves their ability to identify critical test cases. Overall, this helps enhance fault detection for AVs, which is critical for ultimately contributing to the development of safer and more reliable autonomous
systems.
\end{custombox}

\begin{table}[]
    \centering
    \caption{Performance of ML Classifiers}
    \begin{tabular}{c c | c c c}
    \toprule
    \multicolumn{2}{c}{}& \textbf{Dynamic} & \textbf{Static} & \textbf{All}\\
    \midrule
    & P & 0.763 & 0.911 & 0.958\\
    RF & R & 0.805 & 0.869 & 0.915  \\
     & F1 & 0.783 &0.889 & 0.936\\
     \midrule
     & P & 0.723 & 0.860 & 0.897\\
    DT & R & 0.658 & 0.838 & 0.902  \\
     & F1 & 0.689 &0.849 & 0.899\\
     \midrule
     & P & 0.754 & 0.892 & 0.946\\
    KNN & R & 0.766 & 0.849 & 0.867  \\
     & F1 & 0.760 &0.870 & 0.905\\
     \midrule
     & P & 0.770 & 0.928 & 0.942\\
    MLP & R & 0.796 & 0.846 & 0.911  \\
     & F1 & 0.783 &0.885 & 0.927\\
     \midrule
     & P & 0.719 & 0.817 & 0.880\\
    NB & R & 0.730 & 0.730 & 0.818  \\
     & F1 & 0.725 &0.771 & 0.848\\
     \bottomrule
    \end{tabular}
    \label{tab:ml_performance}
\end{table}

\section{Discussion}
\label{sec:discussion}

The features selected by ISA, which includes both static and dynamic aspects, provide critical insights into the road geometry and AV behaviors associated with effective test scenarios. Individually, a single feature is insufficient to fully characterize the road geometry or AV actions that contribute to an effective test case; instead, it is the combination of features that paints a complete picture. For example, the feature \texttt{max\_angle} indicates the maximum angle of the road but does not clarify whether the road is a straight path with a sharp turn at the end or a continuously curved road with a gradual bend. When paired with other road-related features, however, it becomes possible to visualize the types of roads that lead to effective test scenarios. Similarly, features describing AV behaviors, such as \texttt{steering\_min}, reveal the minimum steering value during a test scenario but do not distinguish whether this was due to a straight road or a straight segment within a curved road. Likewise, \texttt{steering\_input\_max} alone cannot determine if the maximum steering input resulted from a sharp turn on a straight road or a continuously curved road. Only by incorporating additional features, such as \texttt{steering\_std}, can we accurately characterize the driving behaviors that correlate with effective test cases. This underscores the importance of analyzing features in combination rather than in isolation.

\begin{figure*}
    \begin{subfigure}[t]{0.3\textwidth}
        \includegraphics[width=\textwidth]{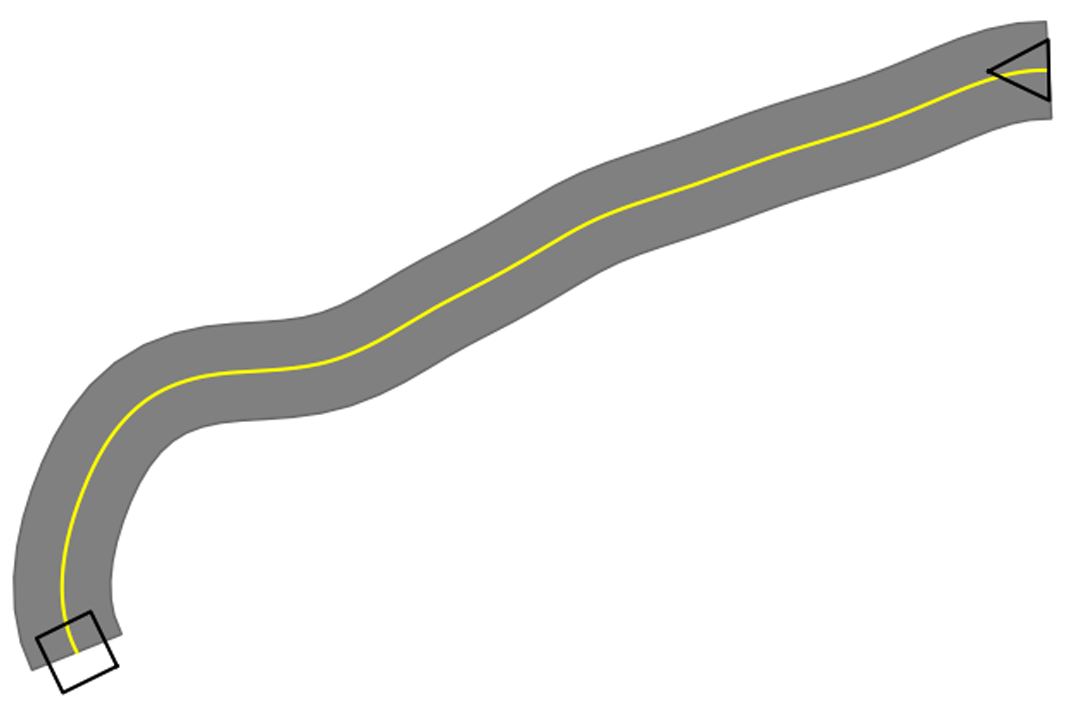}
        \caption{}
        \label{fig:effective_roads-a}
    \end{subfigure}
    \hfill
    \begin{subfigure}[t]{0.3\textwidth}
        \includegraphics[width=\textwidth]{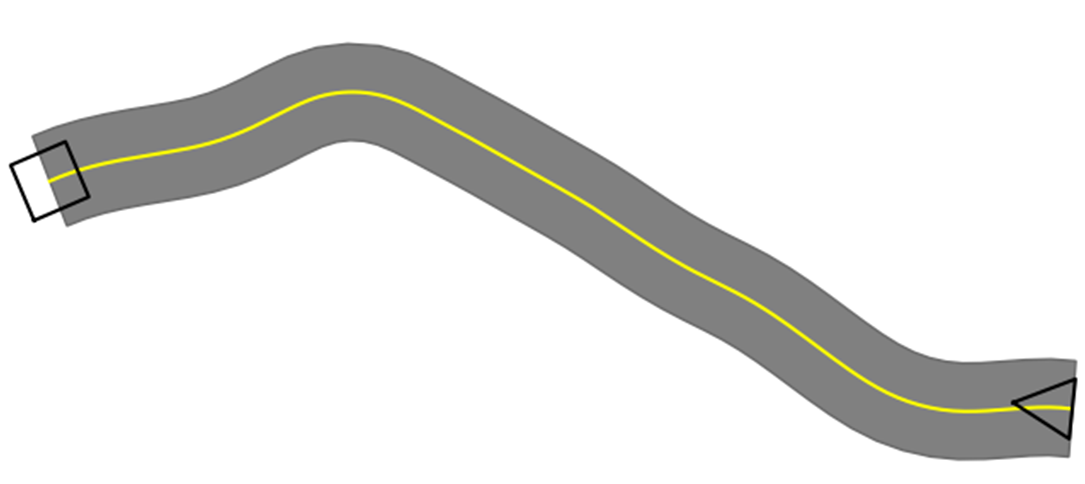}
        \caption{}
        \label{fig:effective_roads-b}
    \end{subfigure}
    \hfill
    % \label{fig:enter-label}
    \begin{subfigure}[t]{0.3\textwidth}
        \includegraphics[width=\textwidth]{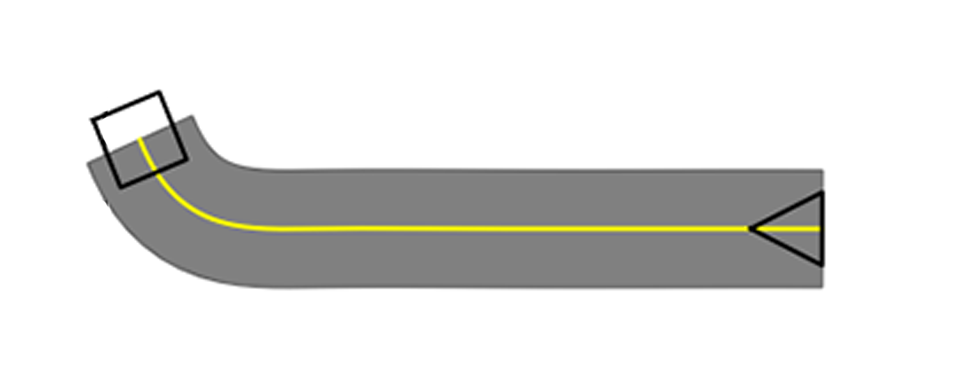}
        \caption{}
        \label{fig:effective_roads-c}
    \end{subfigure}
    \caption{Effective Roads}
    \label{fig:effective_roads}
\end{figure*}

\begin{figure*}
    \begin{subfigure}[t]{0.3\textwidth}
        \includegraphics[width=\textwidth]{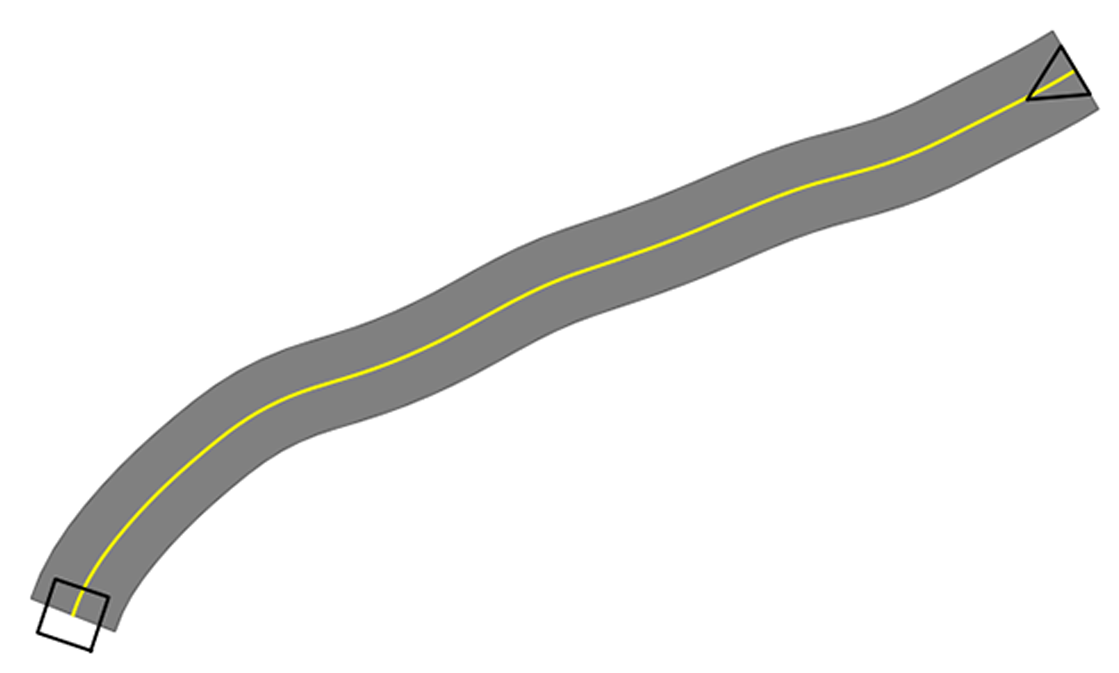}
        \caption{}
        \label{fig:ineffective_roads-a}
    \end{subfigure}
    \hfill
    \begin{subfigure}[t]{0.3\textwidth}
        \includegraphics[width=\textwidth]{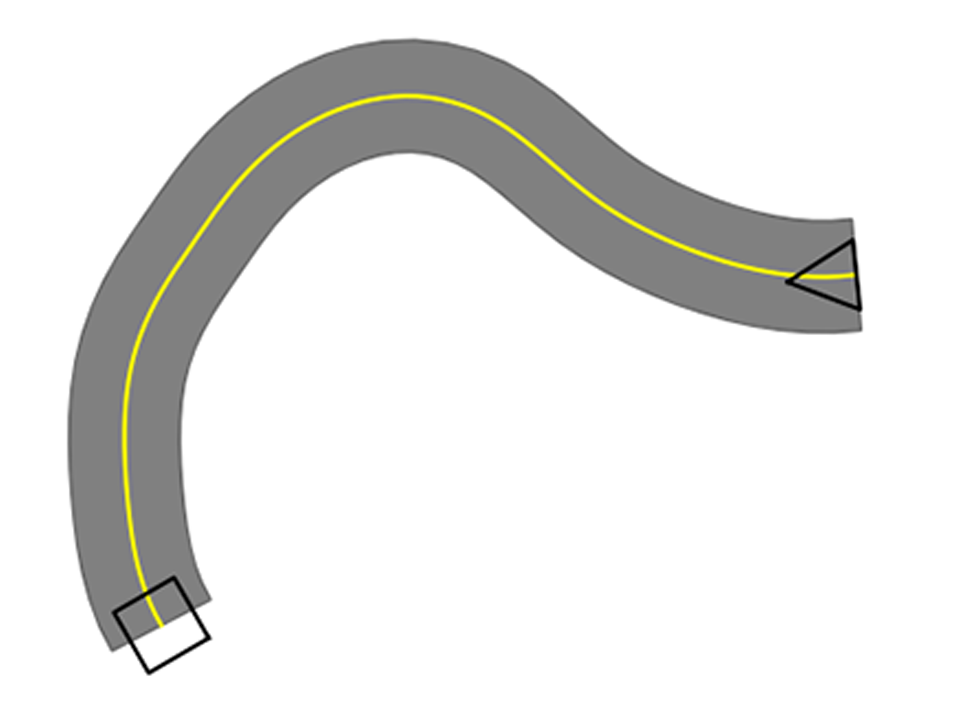}
        \caption{}
        \label{fig:ineffective_roads-b}
    \end{subfigure}
    \hfill
    % \label{fig:enter-label}
    \begin{subfigure}[t]{0.3\textwidth}
        \includegraphics[width=\textwidth]{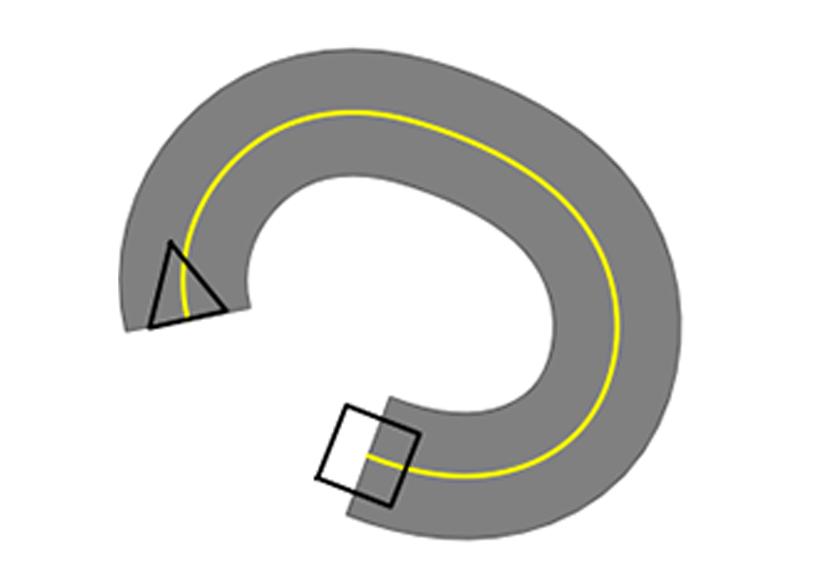}
        \caption{}
        \label{fig:ineffective_roads-c}
    \end{subfigure}
    \caption{Ineffective Roads}
    \label{fig:ineffective_roads}
\end{figure*}

Certain road types, especially those with complex geometry or sharp turns, are more likely to produce effective test scenarios. Roads with straight segments transitioning into abrupt curves challenge the steering, stability, and braking systems of AVs. Such scenarios often require rapid adjustments, testing the AVs responsiveness and control in high-risk conditions, and increasing the likelihood of OOB incidents. Figure \ref{fig:effective_roads-a} displays a road with mostly straight starting segments and gentle curves that do not disrupt the AV operation. In Figure \ref{fig:effective_roads-b}, the road begins with a right turn that the AV successfully navigates, followed by a mostly straight section leading into a curve at the end, which results in an OOB incident. Finally, Figure \ref{fig:effective_roads-c} shows a fully straight road with a sharp turn at the end. 
Each road presents unique challenges for the AV's Lane-Keeping System (LKS), testing its ability to maintain lane position in complex scenarios. LKS failures may indicate a need for controller adjustments or reveal limitations in handling these road conditions. In contrast, Figure \ref{fig:ineffective_roads} presents roads that do not pose a significant challenge for the AV. Figure \ref{fig:ineffective_roads-a} shows a predominantly straight road requiring minimal steering adjustments. Figure \ref{fig:ineffective_roads-b} depicts a road with consecutive curves, and Figure \ref{fig:ineffective_roads-c} features a continuous curve. The geometry of these roads minimizes the AV needing to make sharp steering corrections or accelerate, limiting the potential for testing the AV under challenging conditions.

The static features selected by ISA offer valuable insights into their impact on the AV, as reflected in dynamic features. However, in more complex scenarios with pedestrians or other vehicles, interactions arise among a wider range of features, with these additional actors also influenced by static features and, in turn, impacting AV behavior. Although these interactions in multi-actor environments extend beyond the scope of this study, they represent an intriguing area for future exploration. Another application lies in generating or prioritizing new test cases predicted to be effective. Using predictive models like the machine-learning classifiers in this study, researchers can pinpoint scenarios likely to yield critical insights into AV performance. This approach enables more efficient allocation of testing resources, focusing on scenarios with high potential to reveal unsafe AV behavior, an area we aim to pursue in future research.

The approach presented in this work represents a substantial advancement by integrating insights of what makes a test case effective into the testing workflow. Rather than treating all scenarios as equally valuable, we leverage the specific feature values most strongly associated with effectiveness to guide the testing process. This enables the prioritization of test cases that are statistically more likely to reveal faults, ensuring that the most promising scenarios are executed first and more frequently. This prioritization is particularly valuable in regression testing, where repeated executions can be costly and time-consuming. By ranking and selecting test cases based on the features most predictive of effectiveness, we minimize redundant effort and concentrate computational resources on scenarios with the highest likelihood of exposing safety-critical failures. This approach not only accelerates defect detection but also enhances overall test-suite efficiency.

Beyond immediate efficiency gains, this method transforms feature-based insights into actionable strategies for scenario scheduling, refinement, and long-term testing of AVs. Our framework provides a more targeted, data-driven, and efficient approach to AV testing than previous methods, ensuring that high-impact scenarios are prioritized without compromising coverage or diversity.

\section{Threats to Validity}
\label{sec:threats}

\textbf{Threats to internal validity} may affect the presented results, with one potential threat arising from the selection of features included in the ISA analysis. Feature selection significantly influences the impactful features identified by ISA and can affect the performance of machine learning classifiers. Therefore, the feature selection process is crucial and requires careful consideration. To mitigate this threat, we chose an extensive set of features derived from both static and dynamic aspects of testing test case executions. Similar features have been employed in automated software testing research and AV testing to assess testability and related objectives \cite{AbdessalemNBS18, SDCScissor, birchler2021automated, lu2022learning, cresporodriguez2024isa}.

Another internal validity threat may arise from potential limitations in the cause-effect relationships between scenario-generation technologies and outcomes, influenced by scenario realism, as noted in prior work \cite{DBLP:journals/ese/BirchlerKBGP23}. Our scenarios do not include all real-world elements (\eg weather conditions). However, we used a dataset leveraging different appropriate test subjects (AI drivers), which benefit from robust road knowledge, avoiding issues common to vision-based systems. Future work will utilize new BeamNG features, enabling scenarios with traffic lights, additional vehicles, and static objects.

Similarly, the choice of AV testing techniques and performance metrics is essential in shaping the final results. Adjusting parameters, such as the definition of an effective testing test case during pre-processing, can lead to the selection of different features and alter the outcome of machine learning classifiers, potentially resulting in different outcomes. However, this study does not focus on comparing various configuration parameters or definitions, and interested users can easily replicate the experiments. It is worth noting that this methodology is broad, adaptable, and replicable; final results may vary depending on the selected features, techniques, dataset, and performance metrics.

The choice of training and testing split plays a critical role in evaluating the performance of the machine-learning classifiers. In our experiments, we adopted a standard $80/20$ division of the available test cases, using $80\%$ for training and the remaining $20\%$ for evaluation. While this approach is widely used, it introduces an element of randomness into the pipeline, which can influence classifier performance and lead to variability across runs. Future work will explore alternative partitioning strategies and larger sets of test cases to assess the stability and robustness of the results.

Classifier performance can also depend heavily on hyperparameter configurations. Although we used default settings for this study, fine-tuning these parameters can significantly affect accuracy and generalization. As part of our planned research, we intend to investigate systematic hyperparameter optimization methods, such as grid search or other automated tuning strategies, to identify parameter combinations that yield the best predictive performance for the selected models.

Finally, the control agent governing the AV under test and the simulator used for conducting the tests significantly influence the outcome of the generated instance space. Testing different agents or controllers, such as Lane-Keeping Systems or Autonomous Driving Systems, in different simulators may necessitate different test scenarios, potentially leading to the identification of alternative features and, as a result, a different instance space. Nevertheless, the proposed methodology remains agnostic to the type of control agent or simulator used to create test instances, ensuring its flexibility and ease of application in producing new instance spaces across varying conditions. Similarly, updating or replacing the control agent of the AV or switching to a different simulator for executing test scenarios will equally impact the performance of the machine learning predictors, potentially reducing the accuracy of their predictions. As a result, these predictors will need to be retrained or updated to adapt to the new scenarios, ensuring their continued reliability and effectiveness.

\textbf{Conclusion threats to validity} relate to factors that may impact the accuracy and reliability of the study's conclusions. In our study, we evaluated the effectiveness of AV testing techniques using the commonly adopted performance metric of \textit{Out-of-Bounds incidents}. However, examining alternative metrics, such as safety distance or the frequency of safety constraint violations, and comparing results across multiple metrics could provide similarly valuable insights.

\textbf{Threats to external validity} affect the generalization of the results. To minimize the impact, in our study, we used the dataset generated by  \sensodat \cite{DBLP:journals/corr/abs-2401-09808} which includes three testing techniques used to generate test cases. Such approaches have been widely utilized in previous work on AV testing. Moreover, the testing techniques produce both easy and challenging test cases. Of course, we cannot claim that our results can be generalized to the universe of general open-source AV simulation environments in other domains. Therefore, additional replication studies are desirable, which consider diverse AV data as well as other AV domains.

\section{Related work}
\label{sec:related_work}

This section reviews related work relevant to our investigation such as simulation-based testing, software engineering testing metrics, and regression testing for Cyber-Physical Systems (CPS).

\subsection{Simulation-Based Testing}

A prevalent approach to system-level testing of CPSs, such as unmanned aerial vehicles (UAVs) and self-driving cars (SDCs), involves end-to-end simulation environments where the system under test operates within a simulated physical world~\cite{10195878,DBLP:conf/icst/KhatiriPT23,birchler2021automated}.
However, the applicability and transferability of traditional software testing techniques in these contexts remain uncertain.
New challenges arise in simulation-based testing of CPSs, including simulation realism, computational costs, simulator complexity, and the \emph{Oracle Problem}.
As a result, automated CPS testing continues to be a significant research challenge~\cite{DiSorboTOSEM2023,ZampettiKPP22}.
Simulation-based testing offers a promising approach for improving testing practices in safety-critical systems~\cite{PiazzoniCAYSV21,NguyenHG21,birchler2021automated,DBLP:journals/ese/BirchlerKBGP23,10298301} and supporting test automation~\cite{Wotawa21,afzal2021simulation, wang2021exploratory}.
Previous research has focused on monitoring CPSs and predicting unsafe system states~\cite{DiSorboTOSEM2023,10.1145/3377811.3380353} using simulations~\cite{DBLP:conf/icst/XuAY21,10.1145/3377811.3380353}, as well as programmatic scenario generation~\cite{DBLP:conf/icse/ParkJBC20} or using real-world observations~\cite{DBLP:conf/icst/KhatiriPT23,StoccoPT23}. 

Our study extends these previous studies by introducing a novel approach that combines heuristic-based (\ie feature-based) strategies with ISA to enhance the effectiveness of simulation-based testing while offering insights into the road characteristics and AV behaviors in failing or passing test scenarios.

\subsection{Regression Testing for CPSs}

Simulation-based testing has received considerable attention in recent years, especially in addressing the challenges of test generation for simulation-based tests in areas such as UAVs and AVs~\cite{DBLP:conf/icst/KhatiriPT23, DBLP:conf/iros/ParraO0H23, DBLP:conf/icse/BiagiolaKPR23,GambiJRZ22}.
These methods have yielded significant results in simulation-based testing, promoting advancements in both test generation and optimization~\cite{DBLP:journals/tse/FormicaFRPLM24}.
Moreover, search-based techniques extend beyond test generation, playing a role in regression testing tasks such as test minimization, selection, and prioritization\cite{birchler2021automated,DBLP:conf/splc/ArrietaWSE16,ZHANG20191}.

Arrieta et al. propose a search-based test prioritization method for cyber-physical systems product lines, optimizing five objectives in the fitness function: fault detection capability, coverage of functional and non-functional requirements over time, simulation time, and test case execution time \cite{arrieta2019search}.
In another study, they present a test selection technique for simulation models using the Non-Dominated Sorting Genetic Algorithm-II (NSGA-II), a multi-objective search algorithm \cite{arrieta2018multi}.
Similarly, in \cite{ben2016testing}, a multi-objective search is used to identify safety-critical scenarios in a Pedestrian Detection Vision System for autonomous vehicles, guided by metrics such as car-pedestrian distance, pedestrian proximity to warning zones, and time to collision (TTC) \cite{vogel2003comparison}.
Birchler et al. propose prioritizing test cases for regression testing based on static road features, using a fitness function that favors test cases with greater diversity and lower execution costs, estimated from previous runs \cite{birchler2021automated}.
Lu et al. \cite{lu2021search} develop a test prioritization technique for regression testing, where test case attributes (\eg speed, throttle, weather) are used to compute four properties—diversity, demand, collision probability, and collision information—guiding the search for collision-prone scenarios.

Compared to the aforementioned studies, our approach incorporates insights about what makes a test case an effective one. This knowledge allows us to prioritize the execution of test cases that are more likely to reveal faults. This is particularly beneficial in regression testing, where repeated executions can be costly and time-consuming. By ranking and selecting test cases according to the features most predictive of effectiveness, we reduce redundant effort and concentrate computational resources on scenarios with the highest likelihood of revealing safety-critical failures.

\subsection{Simulation-based Testing Metrics}

Automatically determining the expected test outcome from a given input remains an unresolved challenge, commonly referred to as the oracle problem.
Numerous research efforts have proposed techniques to address this issue in the context of traditional software systems, such as generating oracles~\cite{DBLP:conf/issta/JahangirovaCHT16,DBLP:conf/sigsoft/TerragniJTP20,DBLP:conf/icse/TerragniJTP21}.
Although the oracle problem continues to be an open challenge, requiring human involvement to define the oracle, several metrics, such as code coverage and mutation score, have been introduced to quantitatively evaluate the quality of traditional software systems in support of test automation.
A new area of research has emerged around the automated generation of oracles for testing and fault localization in CPSs using simulation-based technologies.
For instance, Menghi {\em et al.}~\cite{DBLP:conf/sigsoft/MenghiNGB19} introduced SOCRaTes, a method that automatically generates online test oracles in Simulink.
This approach handles CPS Simulink models with continuous behaviors and uncertainties, where oracles are generated from requirements specified in a signal logic-based language.
In this context, as with traditional software testing, simulation-based testing of AVs depends on an oracle to determine whether the observed behavior of a system under test is safe or unsafe.
To support test automation, current research on automated safety assessment of AVs primarily focuses on a set of temporal and non-temporal safety metrics\cite{10.1145/3579642,VR-study,GambiJRZ22,DBLP:journals/ese/BirchlerKBGP23}.
Specifically, the out-of-bound (OOB) non-temporal metric is widely used in simulation-based testing of AVs~\cite{NguyenHG21,GambiJRZ22} to determine whether a test case passes or fails.

\textbf{Instance Space Analysis}.
The Instance Space Analysis (ISA) framework, introduced by Smith-Miles and colleagues~\cite{smith2014towards}, uncovers meaningful relationships between the structural properties of test instances and their influence on algorithm performance.
Originally designed for combinatorial optimization problems, ISA has since been applied across various fields, including automated software testing~\cite{oliveira2018mapping, Neelofar2023}, automated program repair~~\cite{aleti2020apr}, and autonomous vehicle (AV) testing~\cite{neelofar2024identifying, neelofar2024towards}.
In AV testing, ISA helps determine how input features of test scenarios affect their outcomes (safe vs. unsafe) by projecting these scenarios into a 2D space, where the impacts can be visualized effectively.
Neelofar and Aleti~\cite{neelofar2024identifying} created the Instance Space (IS) of simulated test scenarios for AV testing, highlighting the role of scenario features such as pedestrians, road vehicles, weather, and lighting on test results.
Building on this, they proposed a set of test adequacy metrics to assess the diversity and coverage of the test suite \cite{neelofar2024towards}.
Similarly, Crespo-Rodriguez et al. developed the instance space for various search-based test generation techniques aimed at creating challenging road scenarios for AV testing \cite{cresporodriguez2024isa}, providing valuable insights into the strengths and limitations of these techniques. Unlike prior research, which often examines road-based and AV-based features in isolation, this study takes a novel approach by simultaneously analyzing both types of features and investigating their mutual influences. This integrated perspective provides deeper insights into the interplay between static and dynamic factors, a dimension largely unexplored in earlier work. Additionally, this study advances the field by leveraging machine learning classifiers to predict scenario outcomes before execution, systematically comparing models trained exclusively on static features, exclusively on dynamic features, and on a combined set of both. This comparative analysis not only highlights the relative importance of each feature type but also underscores the value of their integration for more accurate and robust predictions.

\section{Conclusion}
\label{sec:conclusion}
This work introduces a technique using Instance Space Analysis (ISA) to identify input features in test scenarios associated with safety-criticality. We examined two feature types: \textit{Static Features}, describing road geometry, and \textit{Dynamic Features}, detailing AV driving operations. Feature distributions enabled a visual assessment of how static features influence AV driving behaviors. Visual representations of test outcomes further allowed us to correlate scenario results with key feature values identified by ISA. Additionally, five machine learning classifiers were trained to categorize scenarios using static, dynamic, and combined features. The classifiers, evaluated by precision, recall, and F1 score, effectively predict test outcomes, potentially reducing the need for simulation and saving testing time.

\section{Data Availability}
\label{sec:data}
Test cases and metadata files, including features and test outcomes are publicly available at \url{https://doi.org/10.5281/zenodo.14919700}. The MATLAB code used for ISA is available at \url{https://github.com/andremun/InstanceSpace} \cite{Smith-Miles_Munoz_2021}.

\bibliographystyle{elsarticle-harv} 
\bibliography{main}

\end{document}